\def\cm{{\mathrm{\thinspace cm}}}

\def\g{{\mathrm{\thinspace g}}}
\def\mg{{\mathrm{\thinspace mg}}}

\def\km{{\mathrm{\thinspace km}}}

\def\m{{\mathrm{\thinspace m}}}

\def\s{{\mathrm{\thinspace s}}}
\def\hr{{\mathrm{\thinspace hr}}}

\def\kmps{\hbox{$\km\s^{-1}\,$}}
\def\kmps{\hbox{$\km\s^{-1}\,$}}

 \documentclass[final,5p,times,twocolumn,authoryear]{elsarticle}

\usepackage{amssymb}

\usepackage{lineno}
\journal{Planetary and Space Science}

\usepackage{url}
\usepackage{amsmath}

\begin{document}

\begin{frontmatter}



\title{Measuring Fluxes of Meteor Showers with the NASA All-Sky Fireball Network}


\author[qualis]{Steven Ehlert}
\author[jacobs,wheaton]{Rhiannon Blaauw Erskine}

\address[qualis]{Qualis Corporation, Jacobs Space Exploration Group, NASA Meteoroid Environment Office, Marshall Space Flight Center, Huntsville, AL, USA, 35812}

\address[jacobs]{Jacobs Space Exploration Group, NASA Meteoroid Environment Office, Marshall Space Flight Center, Huntsville, AL, USA, 35812}

\address[wheaton]{Department of Physics, Wheaton College, 501 College Avenue, Wheaton, IL, USA 60187}

\begin{abstract}
We present an algorithm developed to measure the fluxes of major meteor showers as observed in NASA's All-Sky Fireball Network cameras. Measurements of fluxes from the All-Sky cameras not only improve the Meteoroid Environment Office's (MEO's) ability to provide accurate risk assessments from major showers, but also allows the mass distribution of meteoroids within the shower to be constrained. This algorithm accounts for the shower-specific and event-specific exposure time and collecting area of the sky for nights where sufficiently large samples of shower meteors ($\sim 30$ or more from the shower) are observed. The fluxes derived from the All-Sky Fireball Network for the 2015 Geminid, 2016 Perseid and Quadrantid, 2017 Orionid, and 2018 Leonid shower peaks are calculated. All five of these shower fluxes show excellent agreement with expectations from independent measurements at different mass and luminosity limits. For four of these five showers, the measured mass indices are significantly shallower than what is currently assumed by the NASA Meteoroid Environment Office's (MEO's) annual meteor shower forecast. A direct comparison between forecasted and measured fluxes at limiting masses of $\sim 1 \g$ shows good agreement for the three showers for which the observations took place near their peak activity. 

\end{abstract}
\begin{keyword}

Meteoroids \sep Showers \sep Video




\end{keyword}

\end{frontmatter}

\section{Introduction }
\label{sec:introduction}

The meteor shower forecast and Meteoroid Engineering Model (MEM) produced by NASA's Meteoroid Environment Office (MEO) are fundamentally predictions of the number of meteoroids above a given mass threshold that are expected to impact a spacecraft surface during a particular time. This quantity, known as the flux, is the primary driver of spacecraft risk - a higher flux of meteoroids increases the likelihood of impacts and subsequently risk to the spacecraft \citep[see][for detailed discussions regarding the MEO's shower forecasting and MEM]{Moorhead2019_Forecast,Moorhead2019_MEM}. 

Although flux may be the primary quantity for spacecraft risk, the mass distribution of the meteor shower also plays a role in fully assessing spacecraft risk. Impacts from more massive meteoroids cause more damage to spacecraft. It is therefore imperative to quantify the extent to which a particular meteoroid source may be populated with more massive meteoroids. Such a measurement can be determined using measurements of the flux at different mass limits. Typically, meteoroid masses are assumed to follow a power-law distribution, where the differential number of meteoroids in a given mass bin is given by

\begin{equation}
    \frac{dN}{dm} = N_{0} \left(\frac{m}{m_{\star}}\right)^{-s}
\end{equation}
\noindent where $s$ is known as the mass index, and typically takes on values of $s \sim 1.5-2.5$. The cumulative flux for all meteoroids above a given mass, denoted as $F ( > m)$, is proportional to $m^{1-s}$. We normalize the distribution to some arbitrary characteristic mass $m_{\star}$ in order to maintain clarity in the units of all terms. The normalization constant $N_{0}$ is the differential flux of shower meteoroids at the mass $\m_{\star}$. 

Uncertainties in the mass index can greatly influence the potential risk to spacecraft when extrapolating to high masses. If two sources had the same flux at a mass limit of $1 \mg$ but had mass indexes different by $0.3$ (such as $s=1.7$ and $s=2.0$), the source with the lower mass index would have roughly 8 times the flux of the other at a mass limit of $1 \g$. Since these differences may greatly influence spacecraft design and operations, it is crucial for any model of the meteoroid environment to estimate these mass indices as accurately as possible. 

Fluxes may be an important input quantity for the meteor shower forecast and MEM, but the measurement of meteor shower fluxes requires careful calculations. The MEO has previously measured fluxes only in its Wide-Field camera system \citep{Blaauw2016}, which observes four separate $15^{\circ} \times 20^{\circ}$ regions of the sky for meteors to a limiting mass of $\sim 1 \mg$. The algorithm presented in \cite{Blaauw2016} refines upon previously published algorithms for measuring calibrated shower fluxes \citep[e.g.,][]{Molau2013,Ott2014}, but all of these publications calculate fluxes using a process that assumes a relatively small ($\sim 20^{\circ}-30^{\circ}$) field of view. The flux calculation developed for the Wide-Field cameras does not fundamentally depend on the number of meteors detected, and provides measurements even when as few as one shower meteor has been observed. 

The MEO has not previously measured fluxes from its All-Sky Fireball Network, which would provide an additional measurement of the flux for showers at a mass limit of $\sim 1 \g$ \citep{CampbellBrown2016_MayCam}. This ability would not only provide an independent measure of the flux, but the combined measurements from both camera systems produces a direct measurement of the mass index $s$ across the entire spacecraft threat regime. Meteoroids more massive than $\sim 1 \g$ are generally too rare to have any appreciable chance of impacting a spacecraft. 

The differences in field of view, camera geometry, and number statistics between the All-Sky and Wide-Field systems demand different assumptions and computations. This report will describe how such calculations can be performed for the nights with the highest shower activity using the cameras of the NASA All-Sky Fireball Network. It is crucial to point out that the calculation described in this report can ONLY be performed for meteor showers during their periods of highest activity, as several quantities necessary for calculating a well-calibrated flux will be derived empirically from the observations. These quantities are subject to large uncertainties when the the underlying sample is too small.  

Just as important from the standpoint of spacecraft risk assessment, the process described here will also enable uncertainties on the All-Sky flux and any measurements derived from this flux measurement to be estimated. Throughout this report, all statistical variables are assumed to follow a Gaussian distribution unless otherwise noted, and uncertainties correspond to the central $68.3\% \thinspace (1-\sigma)$ confidence interval.

\section{Needed Quantities and Assumptions}
The measurement of a meteoroid flux from a given source as observed in the All-Sky Fireball Network is defined entirely by four quantities alluded to above. We define them below, and label them with the subscript $i$ when the quantity in question is specific to a single meteor event. 
\begin{enumerate}
    \item The number of meteor events detected that satisfy certain quality cuts, $N$. 
    \item The collecting area of the sky for each event, $A_{i}$.
    \item The total exposure time in which the cameras that detected the event could feasibly observe similar events, $t_{i}$.
    \item The limiting mass/absolute magnitude of the flux measurement, which corresponds to the least massive/luminous meteoroid for which our survey is complete (i.e., where we expect to have observed $\sim 100\%$ of the events that occured over those cameras). We denote the limiting mass and magnitude as $m_{lim}$ and $M_{lim}$, respectively.  
\end{enumerate}
 \noindent The task of any flux measurement is to determine these four quantities as accurately as possible. Whenever relevant, we will compare and contrast the process described here with the Wide-Field algorithm \citep{Blaauw2016}.

\subsection{Assumptions}

The determination of the four quantities described above inevitably requires a large number of simplifying assumptions to be made about the meteor survey performed. Before discussing the All-Sky flux algorithm in detail, we will first explicitly describe the assumptions we make in these calculations. 

\begin{enumerate}
    \item We have a list of meteors observed within an All-Sky camera network. The information calculated for each event includes, at a minimum: the event date/time; its beginning height, its peak absolute magnitude\footnote{Unless otherwise noted, the use of the word magnitude in this manuscript corresponds to the peak absolute magnitude of the meteor observed during its ablation.}, the cameras in the network that detected it, and a shower identification label. 
    \item We can estimate a limiting magnitude for the meteors associated with a given meteor shower on a given night across the entire camera network based on the observed distribution of meteor magnitudes.
    \item For meteors brighter than this limiting magnitude, the meteor survey is complete for all camera combinations and event times. While we expect certain cameras to be able to observe meteors fainter than this network-wide limiting magnitude at certain times, these fainter meteors are not considered in the flux calculation.   
    \item Each meteor event has a specific detector volume defined by the combination of cameras that detected it. 
    \item The detector volume is bound to a range of heights and a minimum elevation angle above the horizon. These same bounds must be satisfied in every camera that detected it. If the meteor event does not satisfy these bounds in a particular camera that detected it, that camera is excluded from consideration in the detector volume calculation. 
    \item The event-specific detector area that goes into the flux calculation is defined as the projection of this volume along the direction of the local radiant at the event time. This projection operation will be discussed in detail in Section \ref{sec:projection}. 
    \item The exposure time associated with a particular meteor event is the total time throughout the night where every camera that detected this event could have feasibly detected other meteors. Note this is an ``intersection'' operation and not a ``union'' operation (i.e., the total clear time in a pair of cameras is the total time where it is clear in BOTH cameras simultaneously).   
    
\end{enumerate}

 With these assumptions in mind, we will now describe the calculations for each of these four quantities (number, area, time, and limiting mass) in detail. 

\subsection{The Meteor Sample}

The flux calculations we are performing will only be performed for meteors that were observed in at least two All-Sky cameras at elevation angles above 30 degrees. This requirement has two purposes: First, it ensures that each meteor has an accurate atmospheric trajectory and orbit determination; and it secondly provides a more accurate estimate of the collecting volume of the sky that is visible to meteoroids. These cuts ensure that the meteoroids curated for each shower and night constitute a pure sample of shower meteors with well defined collecting areas. If a meteor event has been observed by three or more cameras, we consider it to be a detection only in the cameras where this minimum elevation angle requirement is met.

\subsection{Limiting Magnitude}
Because the determination of All-Sky camera fluxes is limited to nights with large samples of shower meteors detected, we can infer the limiting magnitude of the sample directly from the data. The distribution of magnitudes associated with the shower is fit with a Gumbel distribution, following the methodology of \cite{Blaauw2016, Kingery2017}. A Gumbel distribution has a probability distribution function (PDF) of the form 

\begin{equation}
    p(x; \mu, \beta) dx = \frac{1}{\beta}e^{-(z + e^{-z})} dx  \\
\end{equation}
\noindent where the parameter $z$ is defined as $z = \frac{x-\mu}{\beta}$. The parameter $\beta$ serves a similar role to the variance or standard deviation for a Gaussian distribution, and is related to the characteristic width of the distribution near its peak. This choice of PDF has the added convenience that the mode of the PDF is given trivially as $\mu$. We assume that the mode of the best-fit PDF is the limiting magnitude of the shower meteors on that particular night (i.e., $M_{lim} = \mu$). In detail, a histogram with 10 magnitude bins is produced and the best-fit Gumbel distribution parameters are calculated using a least-squares algorithm. This methodology also enables the calculation of the parameter covariance matrix and uncertainties on the model parameters, in particular the mode $\mu$.

\begin{figure*}
    \centering
    \includegraphics[width=0.40\textwidth]{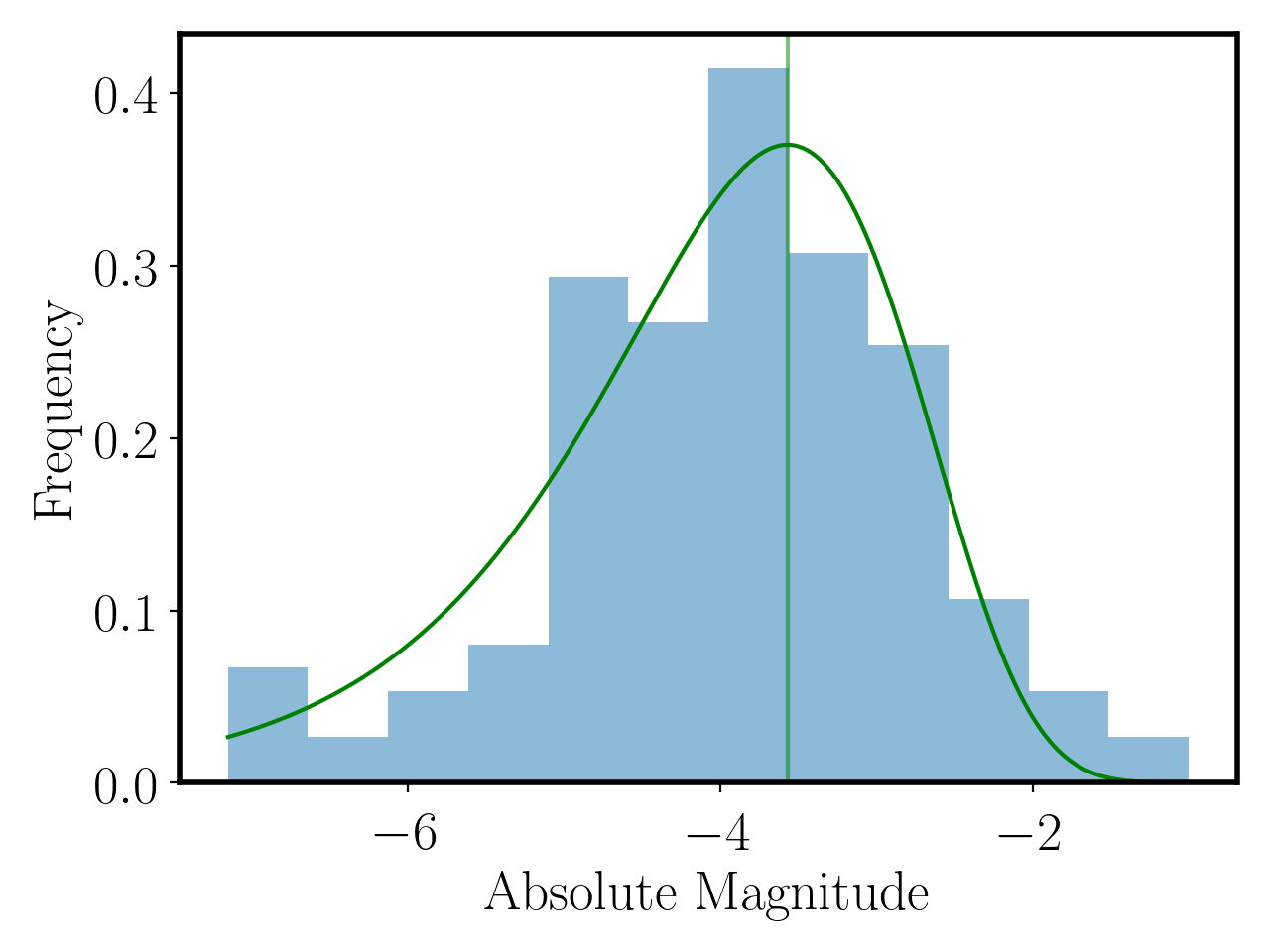}
    \includegraphics[width=0.40\textwidth]{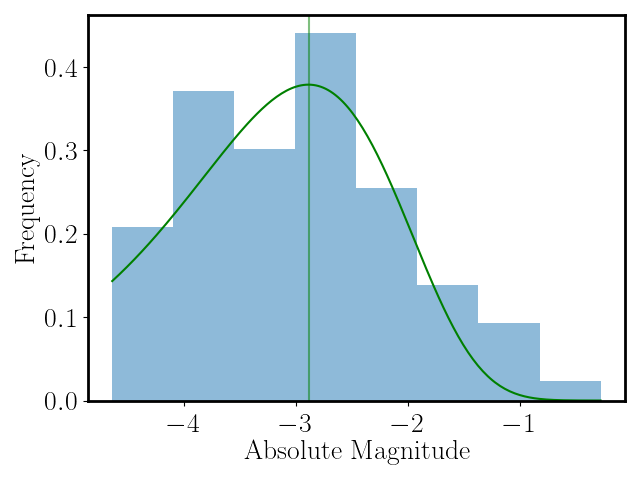}
    \caption{The distribution of peak absolute magnitudes for the Perseid and Geminind meteor showers. In both figures the  green curve denotes the best-fit Gumbel distribution to these histograms, and the vertical green lines denote the limiting magnitude $M_{lim}=\mu$. \textit{Left: } The absolute magnitude distribution for 146 Perseids observed on the night of 2016-08-12, with a limiting magnitude of $M_{lim} = -3.57 \pm 0.10$ and $\beta = 0.99$. \textit{Right: } The absolute magnitude distribution for 79 Geminids observed on the night of 2015-12-15, with a limiting magnitude of $M_{lim} = -2.89 \pm 0.15$ and $\beta = 0.97$. }
    \label{fig:AbsMags}
\end{figure*}

   As stated above we assume that our sample of meteors is complete for magnitudes brighter than the limiting magnitude irrespective of the particular cameras that observed it or the time of observation. The distribution of peak absolute magnitudes for Perseid and Geminid meteors observed on the night of 2016-08-12 and 2015-12-15, respectively, are shown in Figure \ref{fig:AbsMags}. These figures also show the best-fit Gumbel distributions to these data. Only one limiting magnitude is determined for the entire observation network during that night - effectively averaging over the selection functions for each camera and observation time. 

The Wide-Field fluxes algorithm has enough stars in the field of view that a limiting \textit{stellar} magnitude can be modeled in a similar fashion whenever the skies are clear. This limiting stellar magnitude is then converted into a limiting meteor absolute magnitude by correcting for the range and angular motion of the meteor over the duration of one video frame. A limiting meteor magnitude from a particular meteor shower can therefore be inferred for the Wide-Field cameras even when no meteors from the shower are detected. There are insufficient stars too sparsely distributed across the field of view to properly infer the limiting stellar magnitude in the All-Sky cameras, and the limiting meteor magnitude must therefore be derived from the distribution of meteor absolute magnitudes itself. This is the primary reason why shower fluxes can only be measured with All-Sky cameras when large samples of shower meteors are detected.

\subsection{Detector Volume}
We assume that the detector volume of the sky associated with the $i^{th}$ meteor, $V_{i}$ , is independent of the time of observation. It does, however, depend on the combination of cameras that observed it and the shower in question. We assume that the detector volume covers a range of heights based on the empirical distribution of beginning heights for the meteors.

\begin{figure*}
    \centering
    \includegraphics[width=0.60\textwidth]{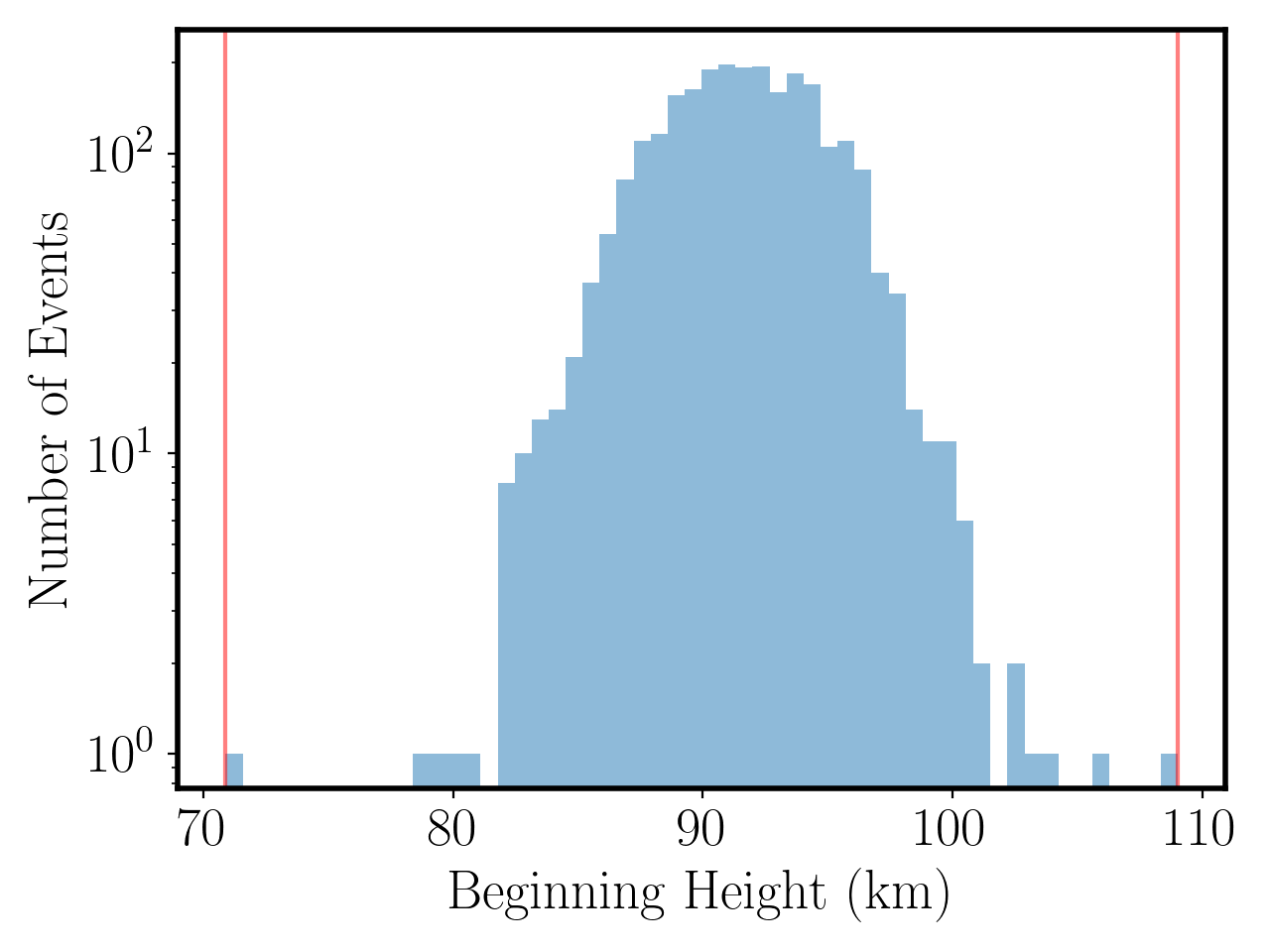}
    \caption{The distribution of beginning heights of all Geminid meteors observed over the lifetime of the NASA All-Sky Fireball Network. The outer red lines correspond to the assumed minimum and maximum heights of the sky detector volume for the Geminids.   }
    \label{fig:BegHeights}
\end{figure*}

 In order to utilize the largest possible range of heights, we set the minimum and maximum heights for the detector volume to the minimum and maximum beginning heights for the entire population of meteors from the shower in question observed by the NASA All-Sky Fireball Network. This includes both meteors observed on the night for which the fluxes are to be calculated as well as any other night where meteors from this shower were observed over the course of the camera network's entire operational lifetime. We restrict the beginning height distribution to only include those events with a convergence angle \citep{Ceplecha_1987_BAIC} of greater than $15^{\circ}$, in order to ensure that the meteor events all have well determined trajectories. The distribution for beginning heights of all observed Geminids are shown in Figure \ref{fig:BegHeights}. Also overlaid on this figure are the assumed minimum and maximum heights for the detector volume. 

With the minimum and maximum heights of the detector volume determined, the full detector volume can be defined. We first use the latitudes and longitudes of every camera along with the minimum and maximum heights to determine a series of points in Earth-Centered Earth-Fixed (ECEF) coordinates. The minimum and maximum values of the ECEF coordinates across all cameras define a large cube that is then discretized into spheres of radius $4 \km$. The choice of spheres simplifies later calculations, and we will discuss that choice in detail later in this report. We identify the positions (in ECEF coordinates) of spheres that are within the previously determined height range and are at an elevation angle above 30 degrees. We only consider volume elements that are at elevation angles above 30 degrees in every camera. This first restriction is to ensure consistency with our elevation angle cutoff for the meteor event sample. This collecting volume is inevitably smaller for events observed in three or more cameras than for events observed only in two cameras, since the collecting volume is defined as the volume at the intersection of all of the camera fields of view. At the end of this calculation, we have the positions of spheres $4 \km $ in radius that satisfy all of these cuts in every All-Sky camera that observed the event. 

The collecting volume of the sky is inherently ambiguous to define due to the absence of clear boundaries and the large range of heights for which meteoroids ablate. There is no fixed ``detector'' in the sky.  Our construction of the detector volume, as described above, was designed to define a common volume for all shower events where they first ``impacted'' it. It is for this reason that beginning heights were chosen, rather than the heights at peak emission. We consider the entire range of beginning heights when calculating the detector volume, as this volume is designed to cover all physically feasible heights over which shower meteoroids could ablate, rather than simply the ``typical'' heights of ablation.   

\subsection{Clear Time}
For the Wide-Field camera network, stack images combining $40 \s$ of frames are produced every 10 minutes. The distribution of pixel values within these stack frames are measured and used to infer whether the image is clear or cloudy. While the same stack images are produced for each All-Sky camera every 30 minutes, the same analysis procedure cannot be applied to these data. Since these cameras have a field of view that covers the entire sky, single value statistics are largely meaningless. Clouds will likely cover some fraction of the field of view while leaving other regions clear. The Moon will certainly be visible in the field of view whenever it is above the horizon. The illumination levels of these features will vary from site to site as well as seasonally within a single site. 

Despite past attempts to develop an algorithmic method of identifying clear and cloudy images in All-Sky cameras, we find that no automated algorithm offers any advantage to visual, manual inspection by a human observer. Since the All-Sky flux algorithm will only be run on nights where there are many meteors detected, the potential concerns of intense and time consuming manual inspection are minimized when the inspection is restricted to these individual nights. Of course, the fact that large samples of meteors were detected on a given night also ensures clear skies were available for at least part of the night. 

 A graphical user interface (GUI) has been developed to automatically find all of the stack images produced by All-Sky cameras during the night in question and keep track of clear and cloudy times identified by the user. The user simply decides whether a particular stack image (and the 30 minutes surrounding it) should be associated with the clear time or cloudy time based on their best-faith guess as to whether or not a meteor could plausibly be detected in the field of view during that time. The limiting magnitude determination takes into account different seeing conditions for different nights.

\subsection{Effective Area}\label{sec:projection}
Unlike the volume of the sky visible to the intersection of All-Sky cameras between two limiting heights, the area of that volume exposed to the shower radiant changes with time. We calculate the effective area for each event, $A_{i}$, by projecting the total ``detector'' volume of the sky along the direction of the radiant. This calculation is greatly simplified by our choice of discretizing the volume into spheres. Since the projected area of a sphere is invariant under rotations ($\pi r^{2}$), the only non-trivial calculation is to determine how many spheres are ``exposed'' to the radiant. 

\begin{figure*} 
    \centering
    \includegraphics[width=0.85\textwidth]{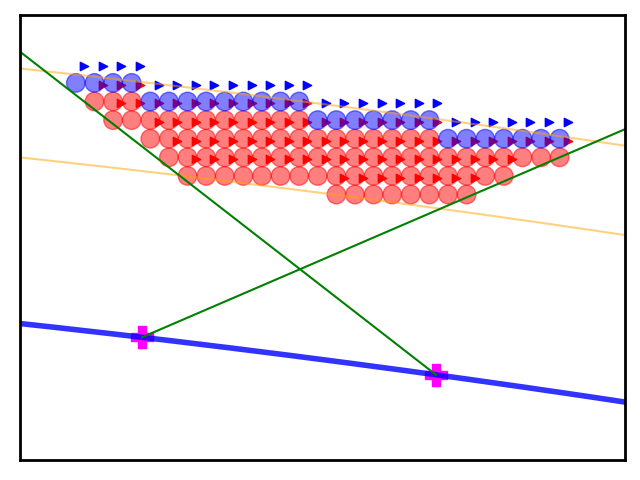}
    \caption{A cartoon depicting the sky area and volume calculations in two dimensions. The positions of the two cameras (the pink crosses) are arbitrarily chosen, but the figure is otherwise drawn to scale for the Earth using the specific choices we assume throughout this work.  The lower and upper orange curves denote altitudes of $71 \km$ and $109 \km$, respectively, similar to what range of heights determined by the histogram shown in Figure \ref{fig:BegHeights}. The green lines denote an elevation angle of $30^{\circ}$ above each camera, which represents our cutoff. The spheres all have radii of $4 \km$, and the triangles denote the test positions for identifying which spheres are exposed to the radiant. Each triangle is located $8 \km$ (1 diameter) away from the center of a sphere in the direction of the assumed radiant. Spheres shaded in blue are exposed to the radiant and contribute to the effective area while spheres shaded in red do not contribute.      }
    \label{fig:CameraGeometry}
\end{figure*}

The algorithm we utilize to determine if a particular sphere is exposed to the radiant is as follows: for each sphere center position within the volume, we move two radii away from that point in the direction of the shower radiant. We then calculate the distance to this new point from every sphere point in the sky volume. If the distance to the closest sphere point is greater than $0.7351$ times the radius of the sphere itself (nominally $4 \km $), then we assume that that sphere point is on the exterior of the volume in the direction of the radiant. Otherwise the sphere point is in the interior of the volume. The particular value of $0.7351$ was determined using Monte Carlo simulations of this grid setup in order to determine the maximum distance between a step of one diameter in a random direction and the nearest sphere point. As a final correction to the collecting area, we multiply the total cross sectional area of all spheres by a factor of $4/\pi$ to account for the area ``missed'' by the spheres\footnote{The ratio of the area of a square to an inscribed circle is $4/\pi$, regardless of the size of the square. This ratio holds even if the square is inscribed by $n \times n$ circles.}. A two-dimensional cartoon of this calculation is shown in Figure \ref{fig:CameraGeometry}. As this figure visualizes, the effective area utilized by this calculation is insensitive to the minimum height of the detector volume, but is far more sensitive to the maximum height.

The Wide-Field fluxes calculation approaches the collecting area determination differently. It finds the area each set of cameras could potentially detect meteors from a specific meteor shower in different height steps, since this area changes significantly across the range of heights meteors are detected. These areas are then corrected for radiant elevation, camera sensitivity, and other selection effects to become the effective area. By considering the entire volume of the sky visible to the radiant at once and determining a collecting area independent of the meteor's height of peak emission, we are able to elegantly account for the spherical geometry of the Earth as well as the potentially complex geometry associated with the regions where multiple cameras intersect.  

Another crucial difference between the Wide-Field flux calculation and this work is that the Wide-Field calculation is able to account for variations in sensitivity across the sky volume and adjust the areas from individual sky elements to account for those variations. We instead estimate a limiting magnitude to which we expect $100\%$ completeness across the entire camera network. No further corrections to the detector area are therefore necessary at these limiting magnitudes. 

\section{The Shower Flux}

The calculations described above provide an event-specific effective area $A_{i}$ and exposure time $t_{i}$. These quantities account for the time independent camera geometry, the shower-dependent beginning heights of their associated meteoroids, and the time dependent camera geometry with respect to the shower radiant. We also account for the times during the night when a meteor could be expected to be detected in every camera. With these corrections already determined, we can define the shower flux over a given time period as 

\begin{equation}
\label{eq:FluxCalc}
    F( < M_{lim}) = \sum_{i} \frac{1}{A_{i} t_{i}}
\end{equation}
\noindent where $A_{i}$ is the effective area for the event accounting for both the geometry of all of the cameras that detected that event and the shower geometry. The effective exposure time $t_{i}$ is the total clear time throughout the night for all of the cameras that observed that event. We emphasize that the total clear time corresponds to an intersection operation and not a union operation - it is the total time throughout the night where the sky was clear over every camera that detected the event.

\subsection{From Magnitudes to Masses}

While the cameras provide a luminosity/absolute magnitude limited flux, the more meaningful quantity from both a physical and engineering standpoint is a mass limited flux. The limiting meteor absolute magnitude is converted into a limiting mass using the model of Brown \citep{Peterson_Spacecraft_Book} to convert the peak magnitude and entry speed of the meteoroid into a mass
\begin{multline}
    2.25 \log_{10}\left(\frac{m_{lim}}{1 \g}\right) =  -8.75 \log_{10}\left(\frac{v}{1 \kmps}\right) \\ - M_{lim} +  11.59
\end{multline}
\noindent which is identical to what was utilized in \cite{Blaauw2016}. This ensures that fluxes with both optical camera systems can be compared directly to one another. This scaling relationship was found in the earlier work of \cite{Jacchia1967} \footnote{Compared to Equation 48 of \cite{Jacchia1967} as written, the formula presented here renormalizes the speeds to $\kmps$ as opposed to $\cm \s^{-1}$ and neglects the weak dependence on zenith angle, effectively assuming a zenith angle of $0$ for all events. Given the weak dependence on zenith angle and the overall uncertainties in mapping magnitudes to masses, this difference is negligible.}, where it was derived using observations of 413 Super-Schmidt meteors along with the classical theory of meteor ablation as discussed in \cite{Verniani1961}. The masses utilized for these Super-Schmidt cameras were derived using the luminous efficiency of \cite{Verniani1964}, which assumes the luminous efficiency depends linearly on speed.

\section{The Equivalent Zenithal Hourly Rate}

In order to compare the measured flux to observations at other sizes/masses, we often compare to other camera systems or different detector instruments. The most consistent source of external data to which we can compare is visual shower counts and Zenithal Hourly Rates (ZHR's). We determine the equivalent ZHR of our measured fluxes at a magnitude of $+6.5$, which is the standard limiting magnitude for visual observations. We assume that the differential distribution of flux per unit luminosity follows a power law in luminosity, $\mathcal{L}$, which we parameterize as 

\begin{equation}
    \frac{dF}{d\mathcal{L}} = N_{0} \left(\frac{\mathcal{L}}{\mathcal{L}_\star}\right)^{-\alpha}
\end{equation}
\noindent for a characteristic luminosity $\mathcal{L}_{\star}$. In this parameterization, the luminosity index $\alpha$ is exactly related to the population index of the shower $r$ as 
\begin{equation}
    \alpha = 1 + 2.5 \log_{10} r
\end{equation}
\noindent This relationship can be extended to a relationship between the population index and mass index through an assumed luminous efficiency. For the luminous efficiency described above, which assumes $\mathcal{L} \propto m^{0.9}$, the mass index is related to the population index as 
\begin{equation}
s=1 + 2.3 \log_{10} r    
\end{equation}

\noindent The ratio of the shower fluxes at two different luminosities is given as 

\begin{equation}
\label{FluxRatio}
    \frac{F( < M_{1})}{F( < M_{2})} = \left(\frac{\mathcal{L}_{1}}{\mathcal{L}_{2}} \right) ^{1-\alpha} 
\end{equation}
\noindent and the luminosity ratio can be derived explicitly from the magnitude system as  
\begin{equation}
\label{LumRatio}
    \left(\frac{\mathcal{L}_{1}}{\mathcal{L}_{2}} \right) = 10^{(M_{1} -M_{2})/-2.5} 
\end{equation}
\noindent Inserting the empirically determined limiting magnitude as $M_{2} = M_{lim}$ and $M_{1} = +6.5$ into Equation \ref{LumRatio} and the resulting luminosity ratio into Equation \ref{FluxRatio} allows the corresponding flux at a limiting magnitude of $+6.5$ to be calculated. This flux can be converted into an equivalent $\mathrm{ZHR}$ for an assumed population index $r$ using the results of \cite{Koschack1990A,Koschack1990B} 

\begin{equation}
    \mathrm{ZHR} = \frac{F (< +6.5) * 37200 \km^{2}}{(13.1 r -16.5)\left(r - 1.3\right)^{0.748}}.
\end{equation}

For this work, the fluxes we measure in the All-Sky cameras are supplemented by flux measurements at other limiting masses using a combination of lunar impact monitoring observations \citep{Suggs2014}, the Wide-Field camera system, and the Canadian Meteor Orbit Radar (CMOR) \citep{Webster2004}. This combination of measurements provides a direct measurement of the mass index $s$. The empirically determined mass index is then used to determine corresponding values of $\alpha$ and $r$, which subsequently determine the equivalent $\mathrm{ZHR}$ value. No independent assumptions about the mass index or flux at a given limiting mass or magnitude are utilized.

\section{Statistical and Systematic Uncertainties}

Just as important as the measurements of the shower flux are the calculations of the expected statistical and systematic fluctuations of that flux measurement. In this section we describe the calculations we perform to estimate these uncertainties. 

\subsection{Limiting Magnitude/Mass Uncertainties}
Since the limiting magnitude is determined using an empirical fit to the meteor sample, the uncertainty in the limiting magnitude determination due to small number statistics cannot be ignored. The particular value of the limiting magnitude affects both the limiting mass and flux of the final measurements. We estimate the uncertainty on the limiting magnitude using the covariance matrix of the Gumbel distribution fit. For the five meteor showers we discuss below, we can fit the limiting magnitude to a precision of $\sim 0.1-0.6 \thinspace \mathrm{mag}$, depending on the size of the underlying sample. In order to account for this uncertainty in our mass index determination, we perform 10,000 Monte Carlo realizations of the limiting magnitude, distributed as a Gaussian with the measured uncertainty, and convert each of these limiting magnitude realization into a limiting mass.  

\subsection{Flux Uncertainties}
There exist two primary sources of uncertainty in the flux measurement as calculated using Equation \ref{eq:FluxCalc} that are necessary to account for.
\begin{enumerate}
    \item In the limit of small numbers of meteors detected that have nearly identical area-time products, the Poisson fluctuations on the total number of shower meteors $N$ will dominate. In this case, the uncertainty on the flux can be estimated as $\sqrt{N} / \bar{At}$, where $\bar{At}$ is the median area-time product for each meteor event considered in the flux calculation.  
    \item In the limit of a large meteor sample with large scatter in their individual area-time products, the variance in these area-time products will be the dominant uncertainty in the flux calculation. For this limit, the uncertainty on the flux is best estimated using bootstrap realizations\footnote{Each of the 10,000 bootstrap realizations utilized is created by resampling the $N$ area-time products with repetition to create new realizations of the data. The distribution of fluxes derived from each of these realizations provides an estimate as to the expected statistical variations we can expect on the measured flux due to the intrinsic scatter in these values.} of the sample of area-time products.

\end{enumerate}

\noindent We assume that these two sources of uncertainty are formally independent of one another. The uncertainty in the measured flux associated with the small number fluctuations, $\sigma_{Pois}$, is calculated as 
\begin{equation}
    \sigma_{Pois} = F / \sqrt{N}
\end{equation}
\noindent which ensures that the \textit{fractional} uncertainty in the flux is equivalent to the \textit{fractional} uncertainty of the underlying sample size, which is itself Poisson distributed. As stated above, bootstrap realizations of the observed sample are utilized to determine the uncertainty associated with the variance in individual event area-time products, $\sigma_{Boot}$. The total uncertainty in the shower flux is given as the sum of these two uncertainties in quadrature, i.e., 

\begin{equation}
    \sigma_{Flux}^{2} = \sigma_{Pois}^{2} + \sigma_{Boot}^{2}
\end{equation}

\subsection{Mass Index Uncertainties}
Mass indices are calculated by fitting the All-Sky fluxes and corresponding fluxes at other limiting masses to a power law model. We determine the uncertainties in the mass index using Monte Carlo simulations that account for the statistical uncertainties in the fluxes at other limiting masses, the uncertainties in the All-Sky flux, and the uncertainties in the limiting mass of the All-Sky flux. 

In determining the mass index using the All-Sky fluxes in conjunction with other flux measurements, there is a crucial caveat to this analysis. The Wide-Field and CMOR derived flux measurements themselves assume a nominal mass index. This mass index is necessarily asssumed to account for the differential sensitivity of each sky element. In order to account for this, we allow the other flux measurements to be distributed over a much larger range than the statistical uncertainties would otherwise suggest. The Monte Carlo simulations sample fluxes that are uniformly distributed over a range of $0.5 \thinspace \mathrm{dex}$\footnote{In other words, for a given flux measurement of $F$, the Monte Carlo simulations sample a range of $[\log_{10}{F} - 0.5, \log_{10}{F} + 0.5]$}, corresponding to a factor of $\sim 3$ in either direction and a fluxes ranging an order of magnitude. This range is also well motivated by the range of physically sensible values of a shower mass index. Since for the other flux calculations the area usually scales as $A^{s-1}$ \citep[e.g.][]{Blaauw_2016_MNRAS} and the nominally assumed mass index is $s=2.0$, a $0.5 \thinspace \mathrm{dex}$ range corresponds to mass index values ranging from $s = 1.5-2.5$. The Monte Carlo simulations vary each flux measurement independently of the others. No such systematic uncertainty is added to the All-Sky flux measurements, as no assumptions of the mass index are necessary for that calculation.

\subsection{Equivalent ZHR Uncertainties}
Using the same underlying Monte Carlo simulations as those utilized for determining the mass index uncertainties, we calculate the distribution of expected equivalent ZHR's given the uncertainties associated with the mass index, the All-Sky flux, and the limiting mass of the All-Sky flux measurement.

\section{Results for Five Showers}\label{sec:ShowerResults}

With the details of this calculation now described, we show the resulting fluxes for All-Sky cameras from several past meteor showers, comparing to other relevant observational data. In particular, the equivalent ZHR's calculated can be compared to observational data from the International Meteor Organization (IMO). 

\subsection{The 2015 Geminid Shower}
The Geminid meteor shower as observed on the night of 14-15 December of 2015 provides an extensive data set for calibrating meteor shower flux measurements. On this night, the MEO not only had robust measurements of the shower activity from the Canadian Meteor Orbit Radar (CMOR) and the Wide-Field cameras, but also observed 33 flashes on the lunar surface over a two hour observation period. Given the nominal rate of lunar impacts of $\sim 0.2-0.5 \hr^{-1}$, the vast majority of these flashes are highly likely to be associated with Geminid shower meteoroids.

\begin{figure*} 
    \centering
    \includegraphics[width=0.55\textwidth]{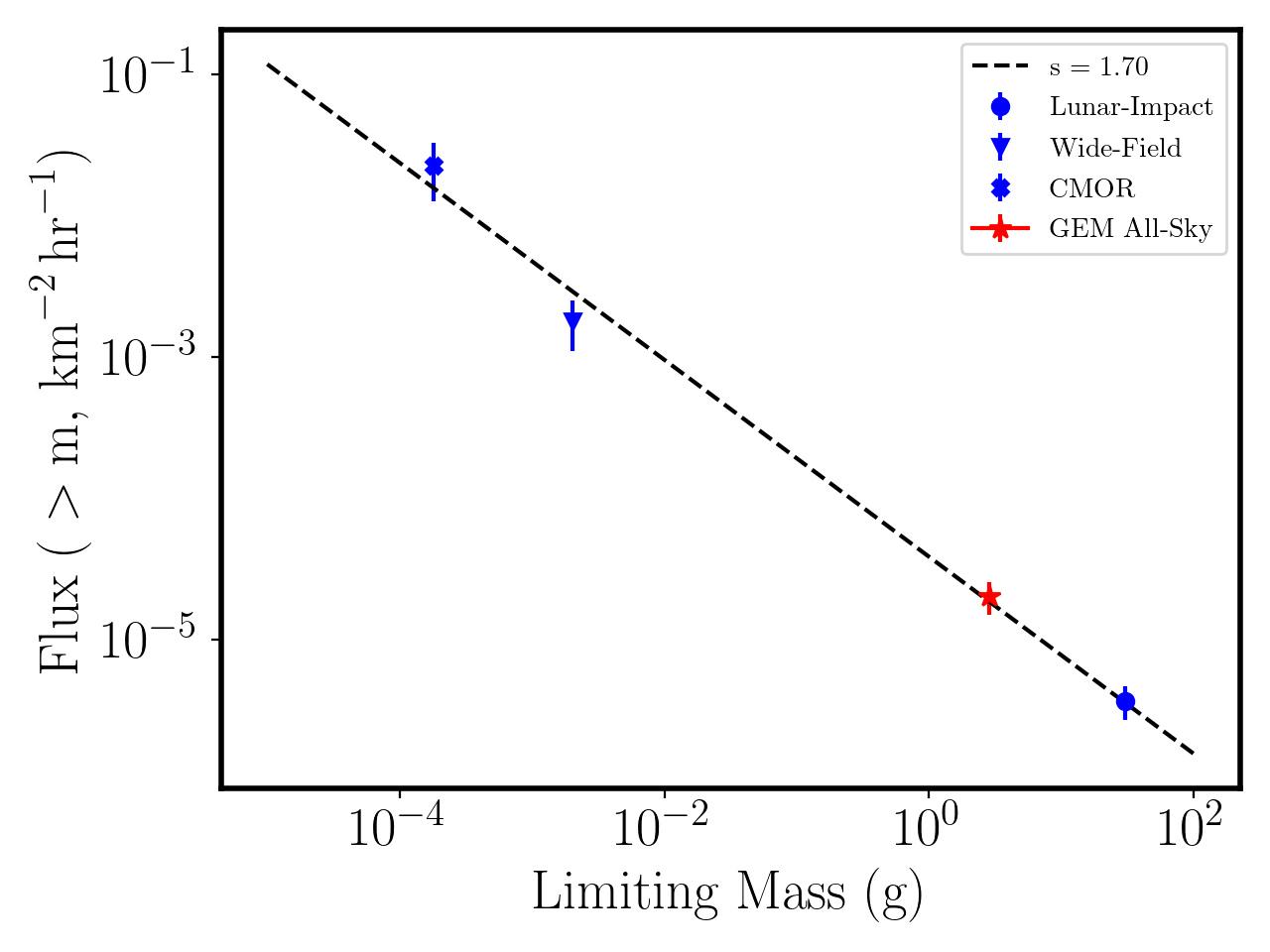}
    \includegraphics[width=0.45\textwidth]{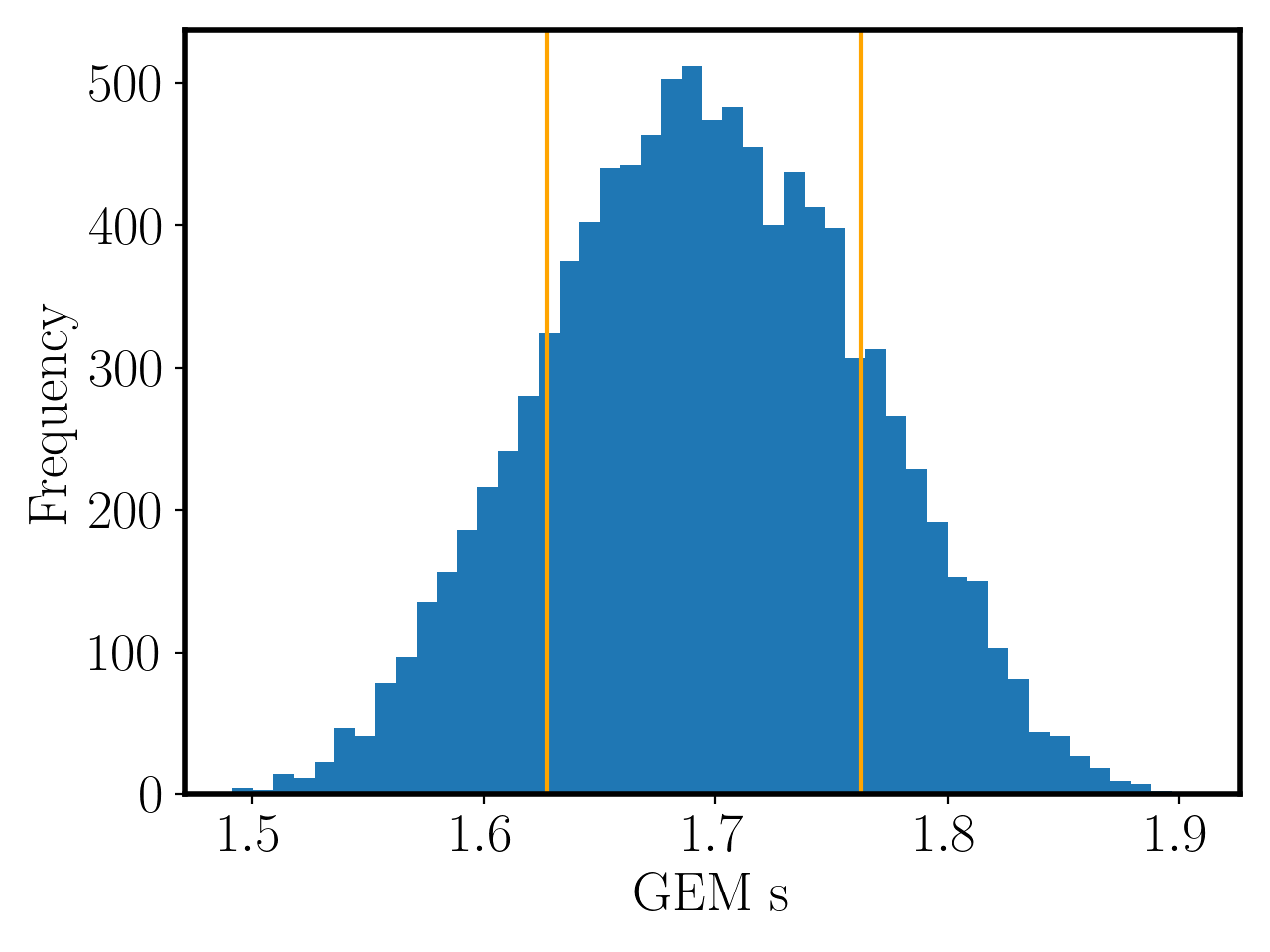}
    \includegraphics[width=0.45\textwidth]{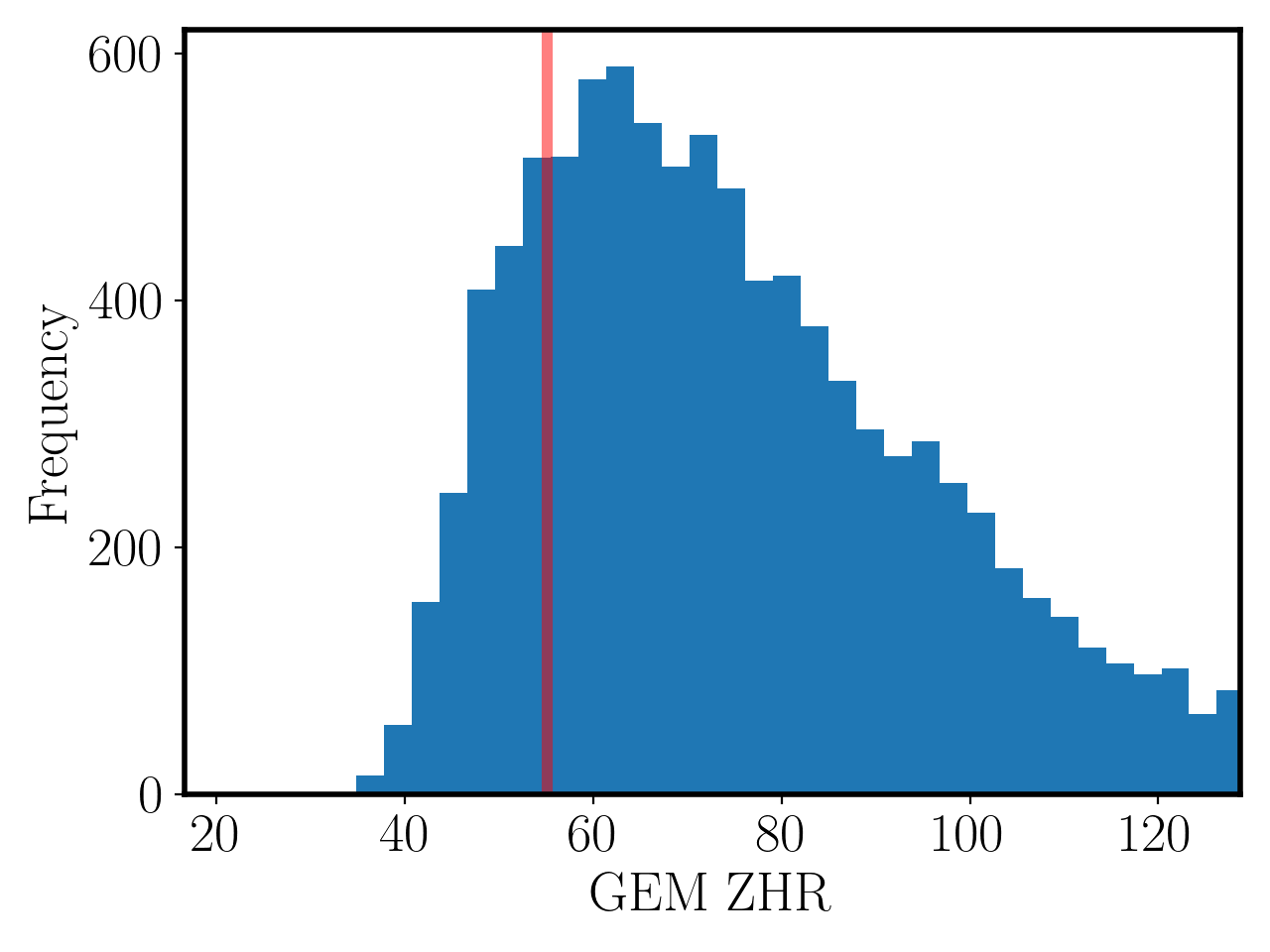}
    \caption{The results for the Geminid shower meteoroids observed on the night of 2015 December 14/15. \textit{Top: } The four flux measurements for the night. The blue markers correspond to the three measurements of \cite{Blaauw2017}, while the red star denotes the corresponding All-Sky flux measurement. The black dashed line denotes the best fit mass index of $s=1.70$. \textit{Bottom Left: } The distribution of Geminid mass indices as determined using Monte Carlo simulations of all four flux measurements and the limiting mass of the All-Sky flux measurement. The vertical orange lines denote the central $68\%$ confidence interval around the best-fit mass index of $s=1.70$. \textit{Bottom Right: } The distribution of equivalent ZHR values at a limiting magnitude of $+6.5$ for the Geminid meteor shower on the night of 2015 December 14/15. The observed visual ZHR is denoted by the vertical red line.    }
    \label{fig:GEM2015}
\end{figure*}

These three instruments provide Geminid shower fluxes at three different limiting masses that span five orders of magnitude in mass: the CMOR limiting mass is $1.8 \times 10^{-4} \g$, while the limiting mass of the lunar impact flashes is $30 \g$. These three flux measurements were combined in \cite{Blaauw2017} to measure the mass index of the Geminid shower as $s=1.70$. For the All-Sky calculation, a total of 93 Geminid meteors were observed across the entire All-Sky camera network. Out of this parent sample, 30 of them were brighter than the limiting absolute magnitude of $-2.89$ and above the minimum elevation angle of 30 degrees. With these values, we determine that the All-Sky camera observations had a limiting mass of $2.85 \g$ and a corresponding flux of $2.04 \times 10^{-5} \km^{-2} \hr^{-1}$. These measurements, plotted in Figure \ref{fig:GEM2015}, are in excellent agreement with the expected flux at this limiting mass using the results of \cite{Blaauw2017}. The resultant mass index measured with all four instruments is $s=1.70 \pm 0.07$. This measured mass index value is in excellent agreement with independent measurements of the Geminid mass index using radar observations \citep{Jones1982,Zigo2009,Blaauw2011,Schult2018} and visual observations \citep{Arlt2006_GEM}. 

Extrapolating the All-Sky flux to the visual magnitude limit of $+6.5$ and this mass index gives an equivalent ZHR of $78 \pm 25 \hr^{-1}$. The average ZHR value for this time window, as determined by visual observers\footnote{Specifically, we used the ten time intervals between 00:00 and 12:00 UT on 2015 December 15 presented in the IMO's Visual Meteor Database \url{https://www.imo.net/members/imo_live_shower?shower=GEM&year=2015}}, was $55 \hr^{-1}$. We reiterate that even for this shower, where we have a prior expectation for the flux within the All-Sky cameras, that the agreement between the equivalent ZHR and that measured by visual observers is not certain. 

\subsection{The 2016 Perseid Shower}
\begin{figure*} 
    \centering
    \includegraphics[width=0.55\textwidth]{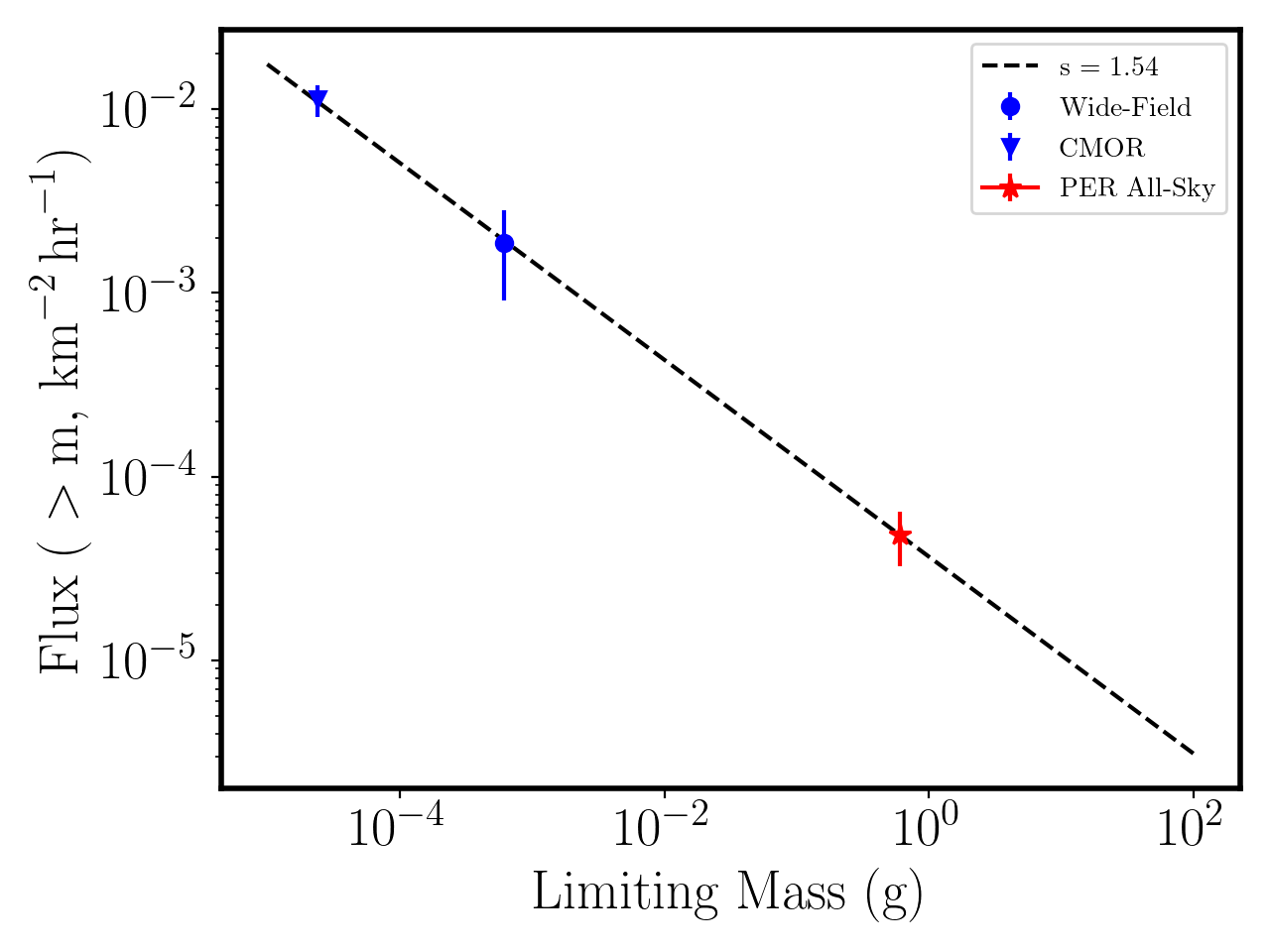}
     \includegraphics[width=0.45\textwidth]{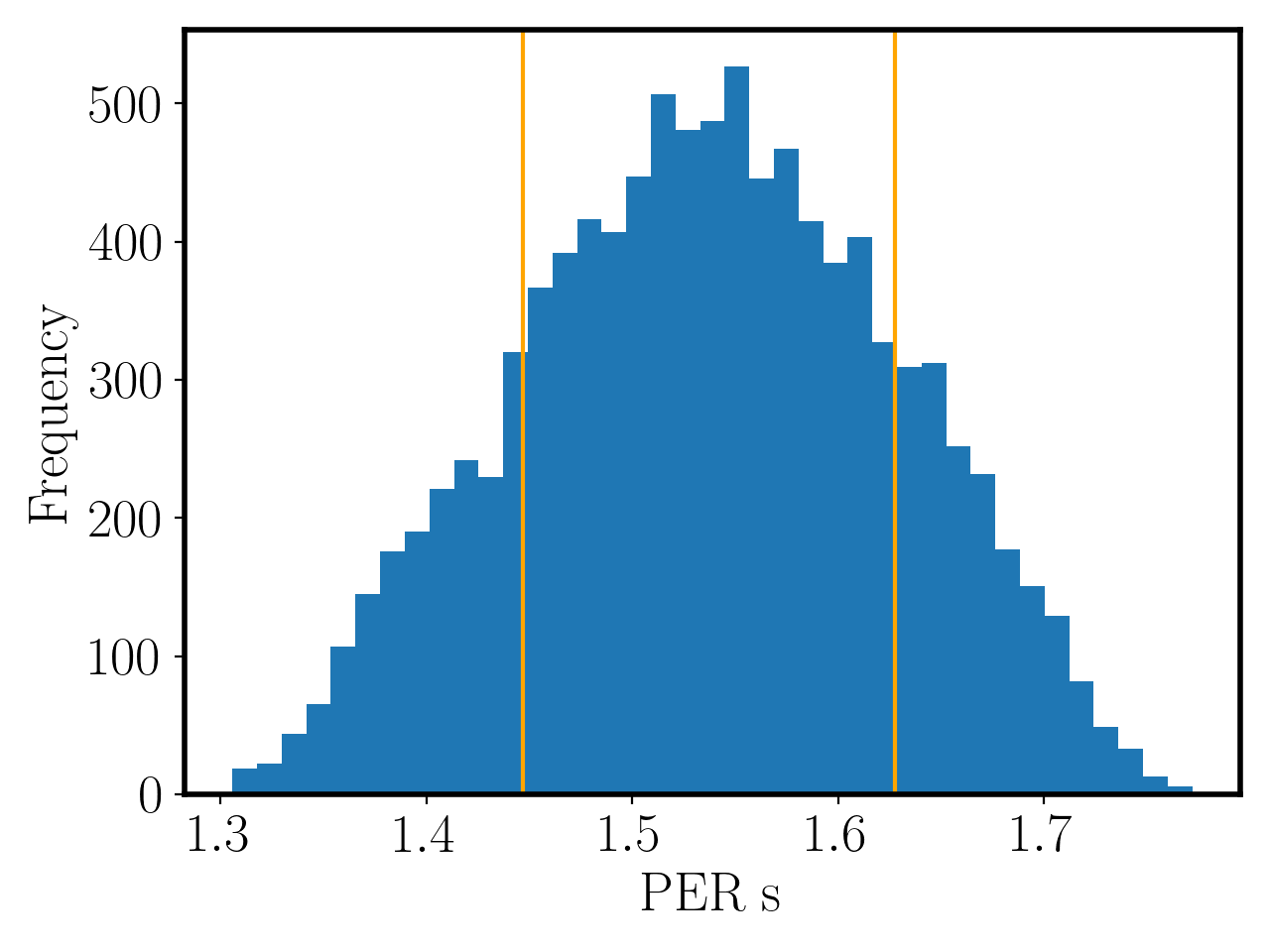}
\includegraphics[width=0.45\textwidth]{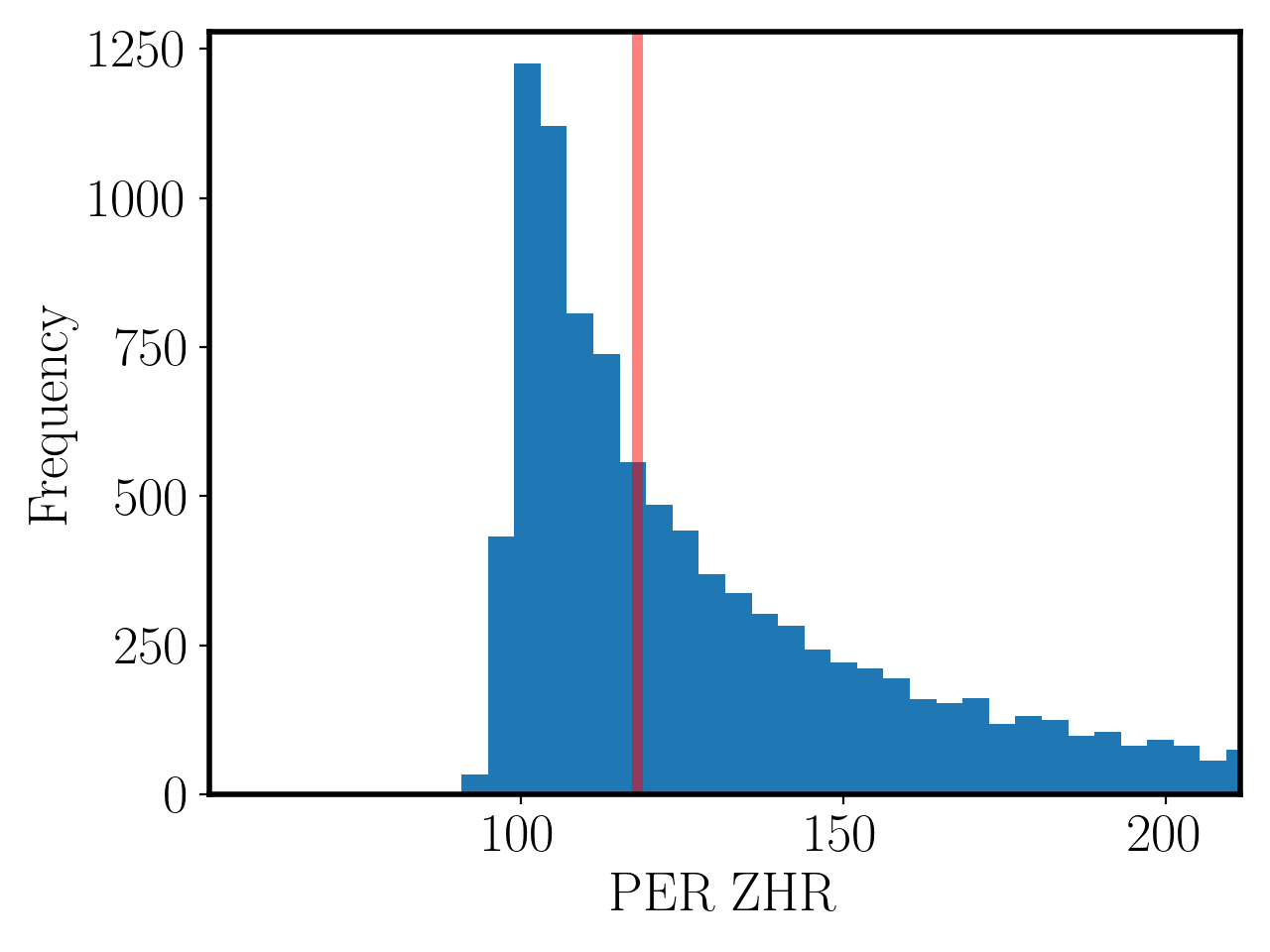}
    \caption{The results for the Perseid shower meteoroids observed on the night of 2016 August 11/12. \textit{Top: } The three flux measurements for the night. The blue markers correspond to the Wide-Field and CMOR flux measurements, while the red star denotes the corresponding All-Sky flux measurement. The black dashed line denotes the best fit mass index of $s=1.54$.  \textit{Bottom Left: } The distribution of Perseid mass indices as determined using Monte Carlo simulations of all four flux measurements and the limiting mass of the All-Sky flux measurement. The vertical orange lines denote the central $68\%$ confidence interval around the best-fit mass index of $s=1.54$.  \textit{Bottom Right: } The distribution of equivalent ZHR values at a limiting magnitude of $+6.5$ for the Perseid meteor shower on the night of 2016 August 11/12. This distribution accounts for uncertainties in the flux as measured in the All-Sky cameras, the limiting mass of the All-Sky camera flux measurement, and the mass index. The observed visual ZHR is denoted by the vertical red line.   }
    \label{fig:PER2016}
\end{figure*}

Our second example shower flux is derived for the Perseid meteor shower on the night of 2016 August 11/12. We combine this measurement with contemporaneous observations in the Wide-Field cameras and CMOR radar system, and the resultant measurements are shown in Figure \ref{fig:PER2016}. A total of 53 Perseid meteors brighter than the limiting magnitude of $-3.34$ were utilized in the flux calculation, derived from a parent sample of 146 Perseid meteors. These numbers result in a flux of $4.86 \times 10^{-5} \km^{-2} \hr^{-1} $ down to a limiting mass of $0.60 \g$. Combining this result with the Wide-Field and CMOR fluxes from the same night gives a mass index of $s=1.54 \pm 0.09$, corresponding to a population index of $r=1.72$. Without the addition of the All-Sky data point (i.e. using only the flux measurements from the Wide-Field cameras and CMOR), the best-fit mass index is $s=1.56 \pm 0.29$. This mass index is shallower than the lowest Perseid mass indices measured in \cite{Hughes1995}, \cite{Jenniskens1998}, and \cite{BrownThesis}, which give typical values of $s \sim 1.7-1.8$. Extrapolating the CMOR flux to the limiting All-Sky mass limit assuming a mass index of $\sim 1.8$ results in a flux that is a factor of several lower than what is measured in the All-Sky cameras, however. The mass index measured for the 2016 Perseids is consistent with previously published telescopic observations of the Perseids in 1969 \citep{Krisciunas1980} and 1992 \citep{Pravek1992}. It is possible that the mass index for this year was genuinely shallower than previous encounters. The shallower mass index we measure here may be, at least in part, due the outbursting nature of the shower during 2016, as a similarly shallow mass index was observed during past Leonid outbursts \citep{BrownThesis}. Other non-outburst and outburst years have had similarly shallow measurements of the mass index \citep{Watanabe1992,Schult2018}. Visual observations measured the population index for the 2016 Perseids to be $r \sim 2.0-2.1$ \citep{Miskotte2017,Molau2017}.

We also compare these fluxes to visual observers by calculating the equivalent ZHR. Extrapolating our flux measurement to visual magnitudes of $+6.5$, spanning four orders of magnitude in luminosity, results in an equivalent ZHR of $134 \pm 39 \hr^{-1}$. A review of archived visual observations shows that the peak ZHR for this night was $180 \hr^{-1}$ and the average visual ZHR during this night\footnote{Specifically, we used the fourteen time intervals between 00:00 and 12:00 UT on 2016 August 12 presented in the IMO's Visual Meteor Database \url{https://www.imo.net/members/imo_live_shower?shower=PER&year=2016}} was $118 \hr^{-1}$. 

\subsection{The 2018 Leonid Shower}
\begin{figure*} 
    \centering
    \includegraphics[width=0.55\textwidth]{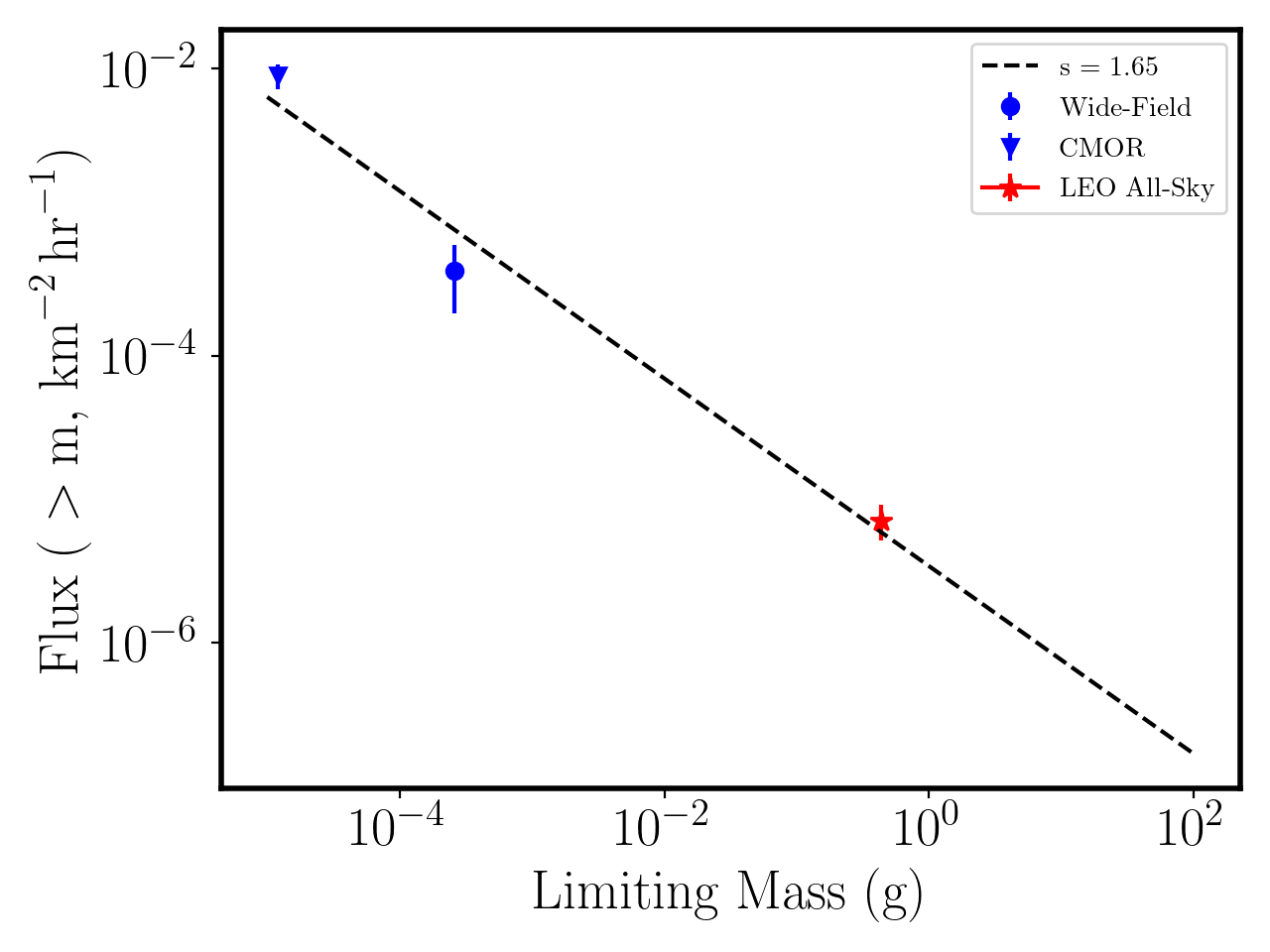}
     \includegraphics[width=0.45\textwidth]{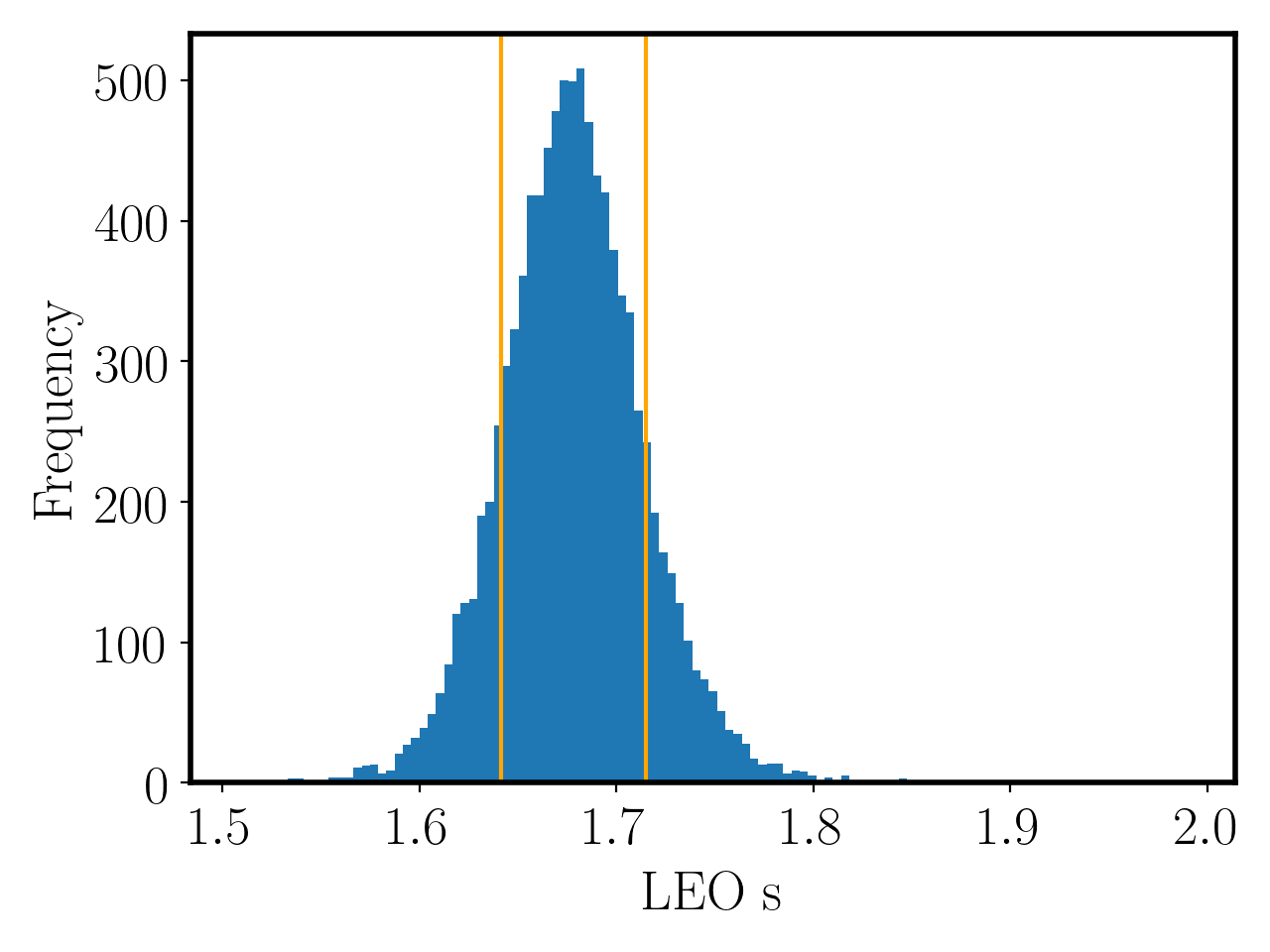}
    \includegraphics[width=0.45\textwidth]{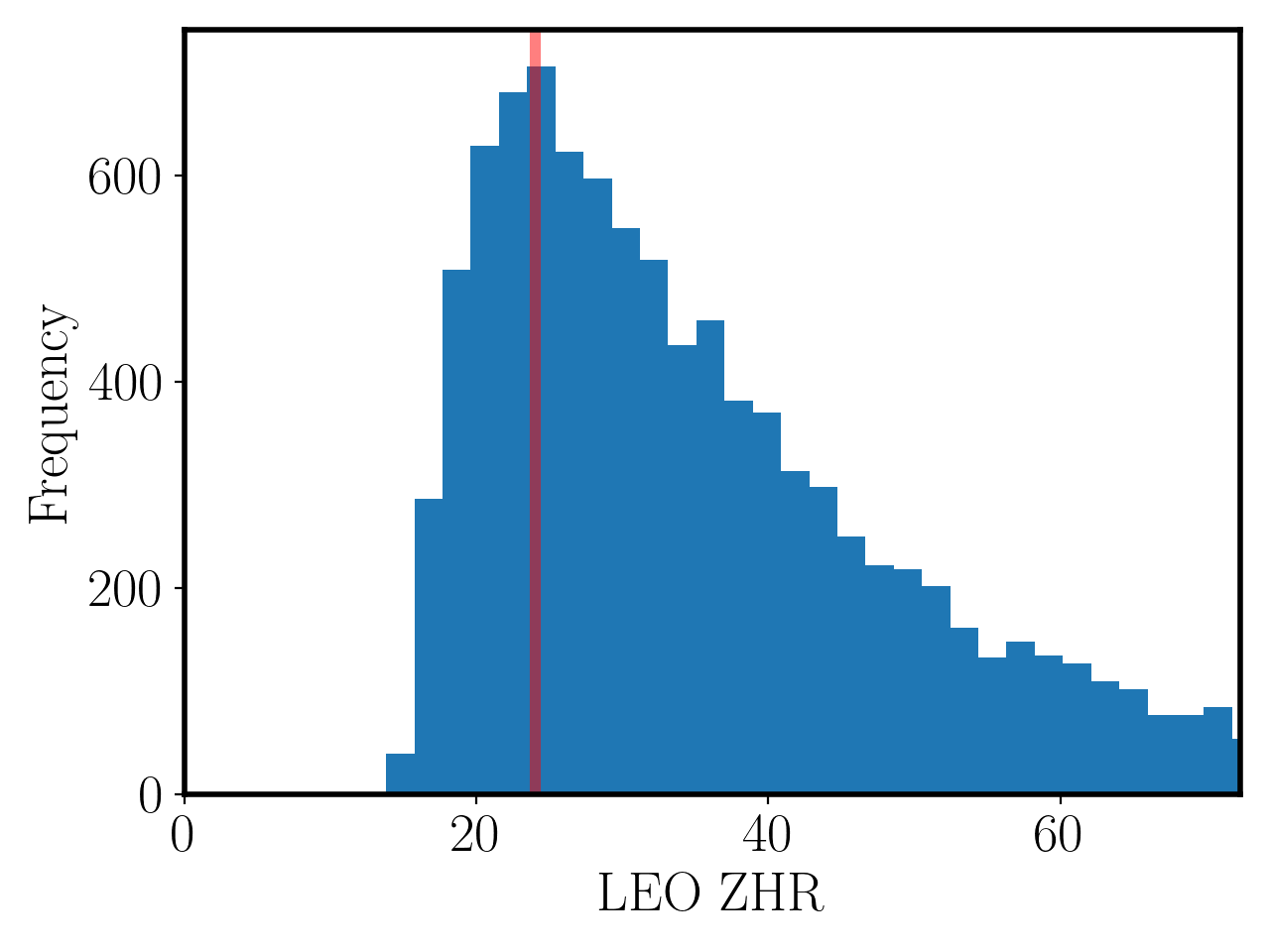}
    \caption{Results for the Leonid shower meteoroids observed on the night of 2018 November 18/19. \textit{Top: } The three flux measurements for this night. The blue markers corresponds to the Wide-Field and CMOR flux measurements, while the red star denotes the corresponding All-Sky flux measurement. The black dashed line denotes the best fit mass index of $s=1.65$. \textit{Bottom Left: } The distribution of Leonid mass indices as determined using Monte Carlo simulations of all three flux measurements and the limiting mass of the All-Sky flux measurement. The vertical orange lines denote the central $68\%$ confidence interval around the best-fit mass index of $s=1.65$. \textit{Bottom Right: } The distribution of equivalent ZHR values at a limiting magnitude of $+6.5$. This distribution accounts for uncertainties in the flux as measured in the All-Sky cameras, the limiting mass of the All-Sky camera flux measurement, and the mass index. The observed visual ZHR from this night is denoted by the vertical red line. }
    \label{fig:LEO2018}

   \end{figure*}

Outside of the Perseid and Geminid showers, the single night where the most meteors were detected by the All-Sky Fireball Network was 2018 November 18/19. A total of 15 Leonid meteors out of an initial sample of 38 were brighter than the limiting magnitude of $-3.61$. The corresponding flux is measured as $7.16 \times 10^{-6} \km^{-2} \hr^{-1}$ down to a limiting mass of $0.43 \g$. The equivalent ZHR of this measurement is $37 \pm 17 \hr^{-1}$, a value that is in excellent agreement with visual observations of the Leonids during the neighboring nights. The resultant mass index of the Leonid shower across all three instruments is $s=1.65 \pm 0.09$, which is a value nearly identical to that of the Geminids. These measurements are all shown in Figure \ref{fig:LEO2018}. This mass index is shallower than that measured by \cite{Koten2011}, but in good agreement with the visually-determined measurement of \cite{Jenniskens2009}.  

\subsection{The 2016 Quadrantid Shower}
\begin{figure*} 
    \centering
    \includegraphics[width=0.55\textwidth]{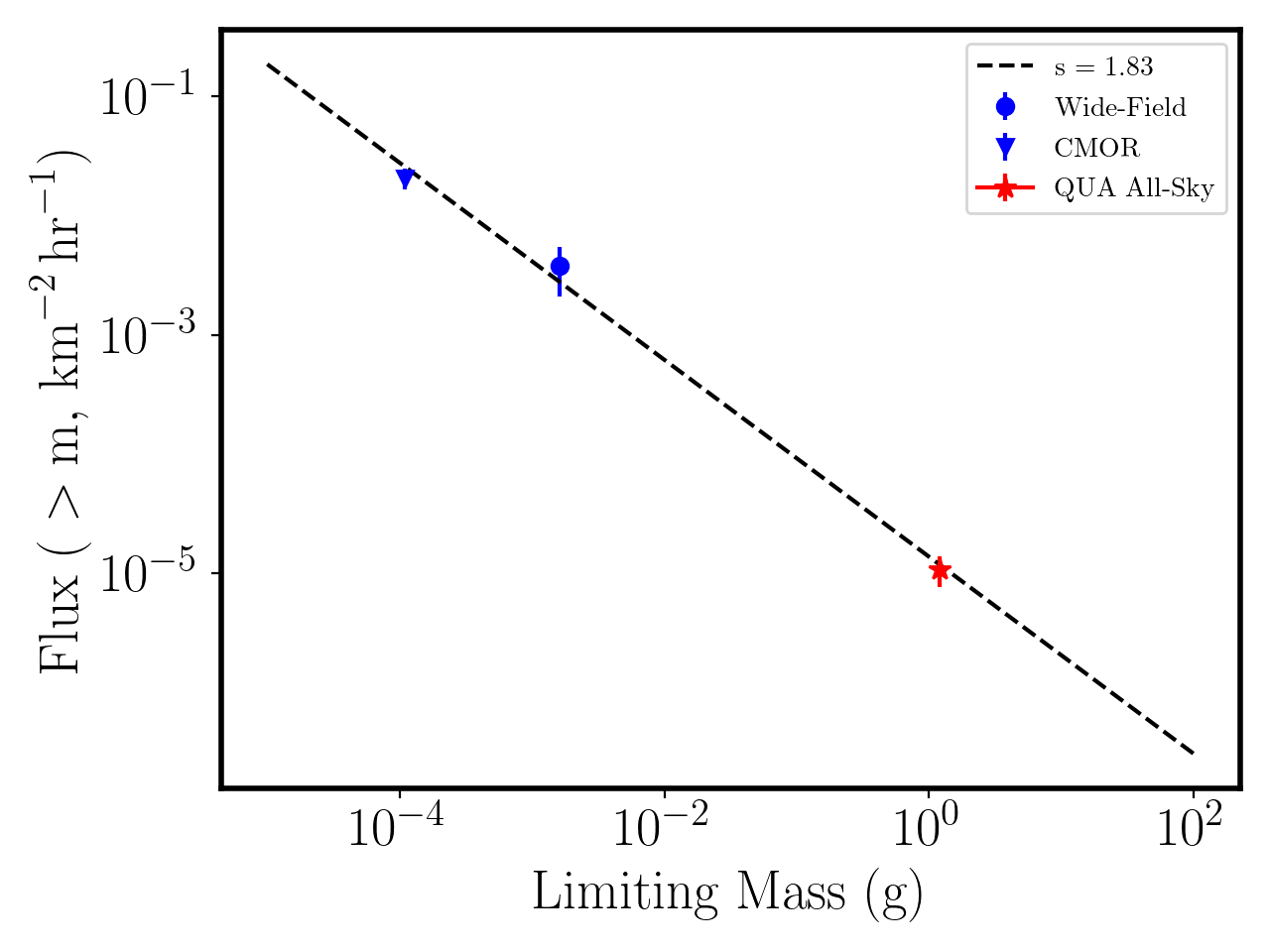}
    \includegraphics[width=0.45\textwidth]{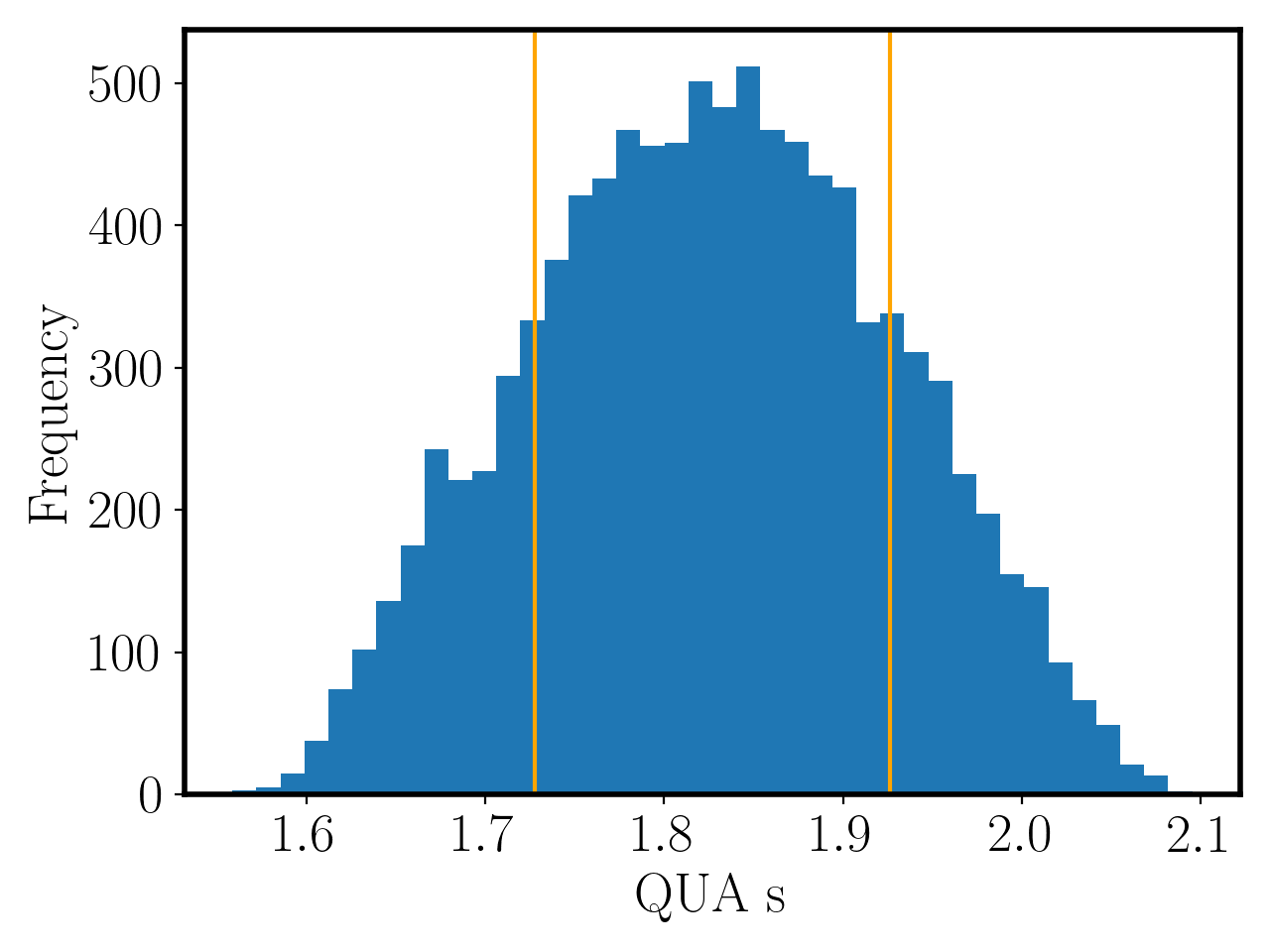}
    \includegraphics[width=0.45\textwidth]{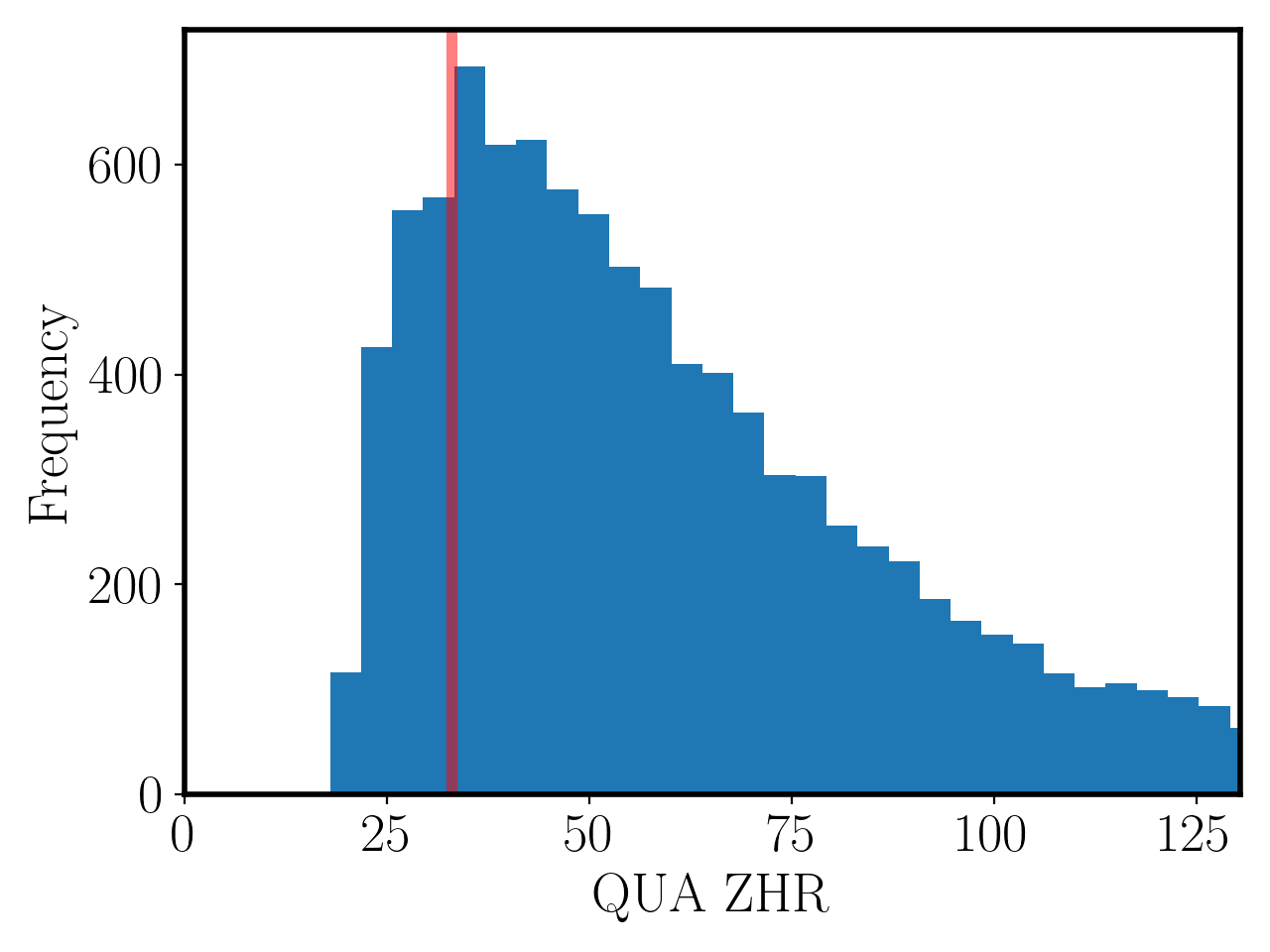}

    \caption{Results for the Quadrantid shower meteoroids observed on the night of 2016 January 3/4. \textit{Top: } The three flux measurements of this night. The blue markers corresponds to the Wide-Field and CMOR flux measurements, while the red star denotes the corresponding All-Sky flux measurement. The black dashed line denotes the best fit mass index of $s=1.83$. \textit{Bottom Left: }  The distribution of Quadrantid mass indices as determined using Monte Carlo simulations of all three flux measurements and the limiting mass of the All-Sky flux measurement. The vertical orange lines denote the central $68\%$ confidence interval around the best-fit mass index of $s=1.83$. \textit{Bottom Right: } The distribution of equivalent ZHR values at a limiting magnitude of $+6.5$ for the Quadrantid meteor shower on the night of 2016 January 3/4. This distribution accounts for uncertainties in the flux as measured in the All-Sky cameras, the limiting mass of the All-Sky camera flux measurement, and the mass index. The observed visual ZHR from 2016 January 3/4 is denoted by the vertical red line.    }
    \label{fig:QUA2016}
\end{figure*}

We next measure the flux for the Quadrantid shower on the UT date of 2016 January 3/4, with a total of 40 Quadrantid meteors detected across all cameras. Out of these events, a total of 15 Quadrantid meteors are brighter than the limiting magnitude of $-2.56$. The corresponding flux was measured as $1.08 \times 10^{-5} \km^{-2} \hr^{-1}$ down to a limiting mass of $1.20 \g$. Figure \ref{fig:QUA2016} shows this flux combined with measurements from the Wide-Field cameras and CMOR radar system, giving a best fit mass index is $s=1.83 \pm 0.10$. The equivalent ZHR of this measurement is $63 \pm 33 \hr^{-1}$, which is in good agreement with visually observed $ZHR = 33 \hr^{-1}$ of the Quadrantids during this night. The mass index of $s=1.83$ measured here is in good agreement with previous measurements in the radar \citep[$s=1.6-2.0$,][]{Browne1957,Blaauw2011} and visual observations \citep[$r \sim 2.1-2.4$,][]{Rendtel1993,Weiland2012}. This mass index is steeper than those observed in \cite{Brown1998,Rendtel2016,Schult2018}, which consider lower limiting masses. 

\subsection{The 2017 Orionid Shower}
\begin{figure*} 
    \centering
    \includegraphics[width=0.55\textwidth]{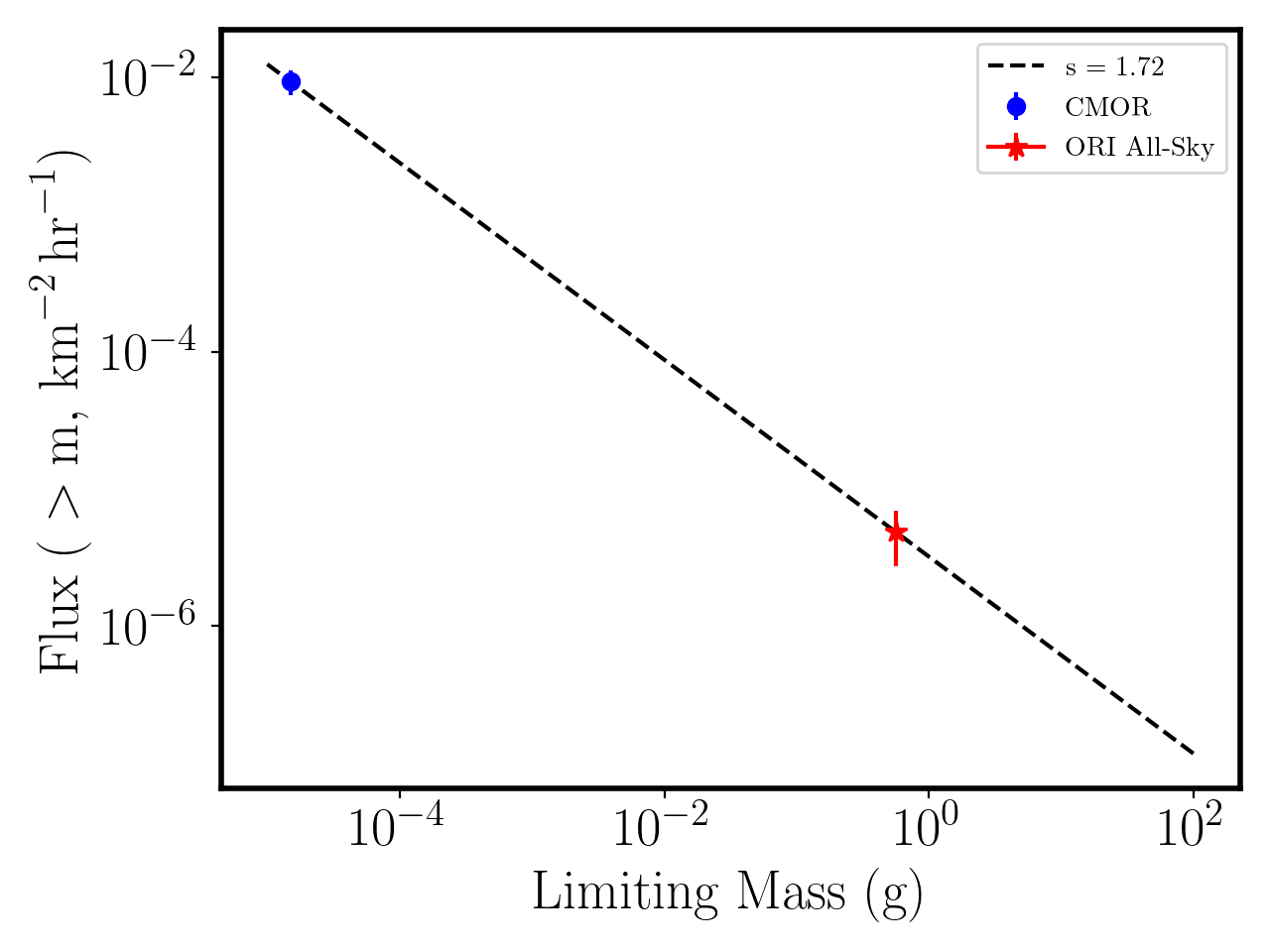}
    \includegraphics[width=0.45\textwidth]{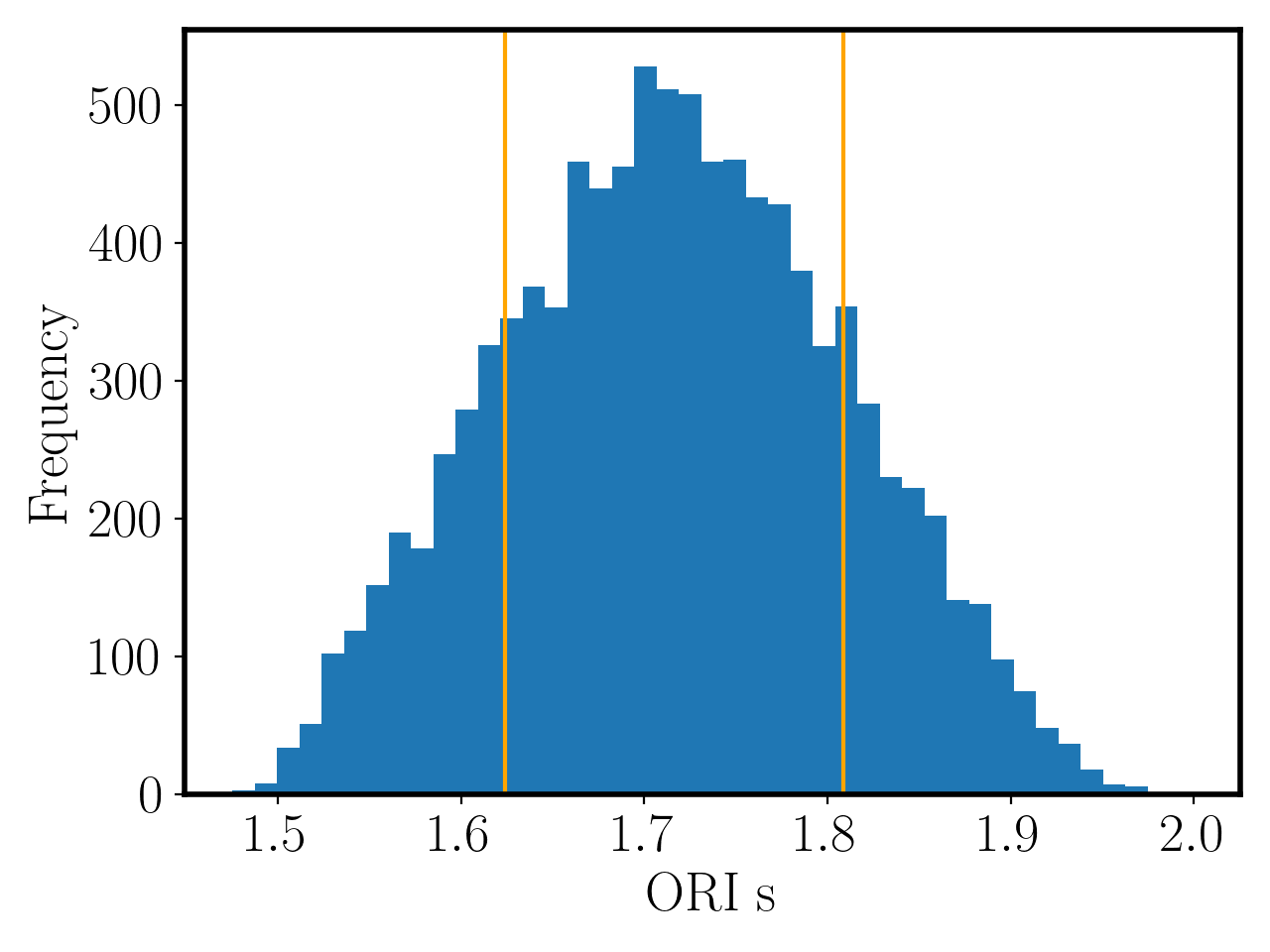}
\includegraphics[width=0.45\textwidth]{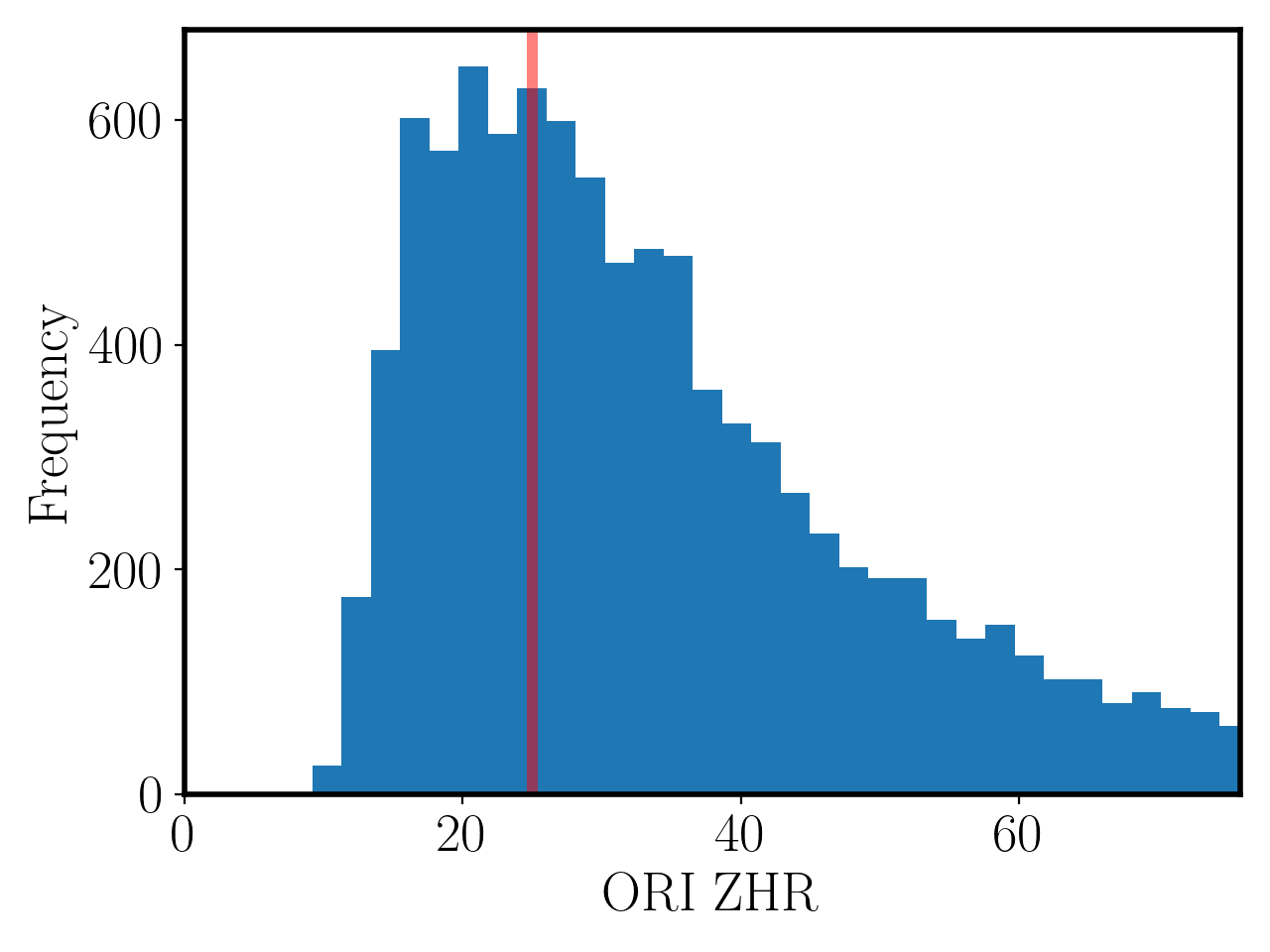}
    \caption{The results for the Orionid shower meteoroids observed on the night of 2017 October 21/22. \textit{Top: } The three flux measurements of this night. The blue circle corresponds to the CMOR flux measurement, while the red star denotes the corresponding All-Sky flux measurement. The black dashed line denotes the best fit mass index of $s=1.72$.  \textit{Bottom Left: } The distribution of Orionid mass indices as determined using Monte Carlo simulations of all three flux measurements and the limiting mass of the All-Sky flux measurement. The vertical orange lines denote the central $68\%$ confidence interval around the best-fit mass index of $s=1.72$. \textit{Bottom Right: } The distribution of equivalent ZHR values at a limiting magnitude of $+6.5$ for the Orionid meteor shower on the night of 2017 October 21/22. This distribution accounts for uncertainties in the flux as measured in the All-Sky cameras, the limiting mass of the All-Sky camera flux measurement, and the mass index. The observed visual ZHR from this same night is denoted by the vertical red line.}
    \label{fig:ORI2017}
\end{figure*}

We finally measure the flux for the Orionid shower on the UT date of 2017 October 21/22, with a total of 28 Orionid meteors detected across all cameras. Out of these events, a total of 9 Orionids were brighter than the limiting magnitude of $-3.64$ were identified. The corresponding flux was measured as $4.85 \times 10^{-6} \km^{-2} \hr^{-1}$ down to a limiting mass of $0.56\g$. Combining this flux with contemporaneous measurements from the CMOR radar system gives a best-fit mass index of $s=1.72 \pm 0.09$, as shown in Figure \ref{fig:ORI2017}. The equivalent ZHR of this measurement is $37  \hr^{-1}$, with large uncertainties. This value is in good agreement with visually observed $ZHR = 25 \hr^{-1}$ during this night. The mass index measured in this work is consistent with the results of \cite{Jones1989} and \cite{Rendtel2008_ORI}, although it is on the low end of values measured by \cite{Blaauw2011}. The radar results of \cite{Schult2018} found a steeper mass index of $s=1.95$.

\section{Systematic Uncertainties }

While the agreement between the various flux measurements for these five showers with the All-Sky flux algorithm presented here is encouraging, there are two main sources of systematic uncertainties/effects that may affect the possible interpretation. In this section we will discuss our efforts to quantify the extent of these effects and justify that the fluxes we present are robust to these systematic effects.

\subsection{Sky Geometry}

We make one choice about the collecting area of the sky in this flux calculation that requires further justification - an elevation angle cutoff of $30^{\circ}$. In this section, we discuss how small changes to this otherwise fixed quantity affects the measured All-Sky fluxes for the Geminid meteor shower, the one shower where we have external prior information for both the mass index and ZHR. While it would be ideal to perform this test with additional showers as well, there exist no independent observations for these showers that can be used to constrain the flux at a limiting mass of $\sim 1 \g$.  

The assumed elevation angle cutoff affects the input meteor sample as well as the collecting area of the calculation. In the ideal case, the measured flux of shower meteors should not depend on the choice of elevation angle cutoff. A stricter elevation angle cutoff should reduce the number of shower meteors by the same factor as the collecting area, leaving the flux invariant. In order to maximize the sample and statistical precision of the measurement, however, the lowest acceptable elevation angle cutoff should be used. 

\begin{figure*} 
    \centering
    \includegraphics[width=0.45\textwidth]{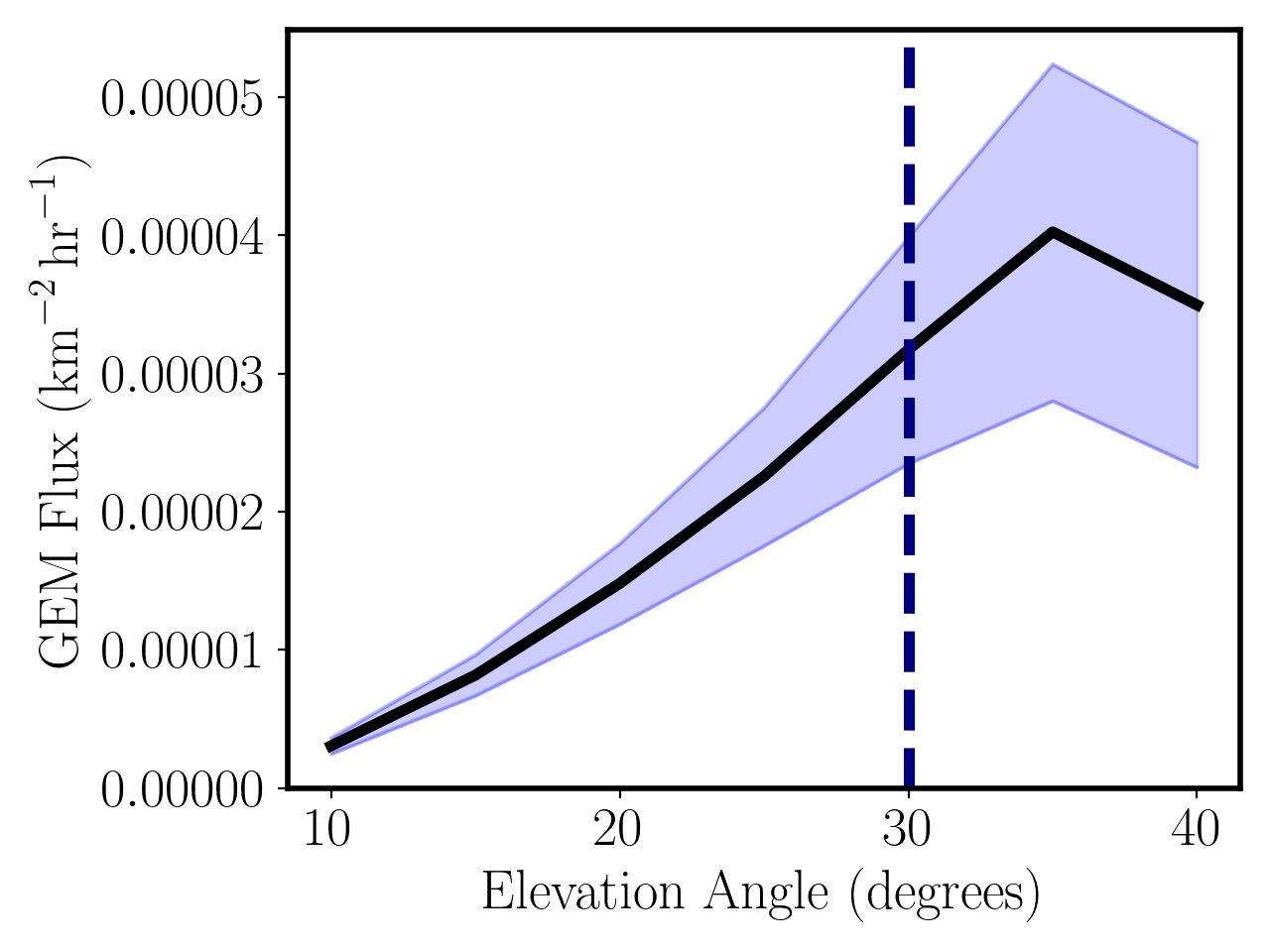}
    \caption{The sensitivity of the measured Geminid shower flux as a function of the minimum elevation angle cutoff. The cutoff chosen for this work, 30 degrees, is denoted by dashed line. For more restrictive cutoffs (above 30 degrees) the flux is consistent with a constant value, whereas a clear decrease in flux is observed for more generous elevation angle cutoffs. This is precisely the trend we expect when the observed sample of meteors at low elevation angles is incomplete.  }
    \label{fig:ElevationAngleFlux}
\end{figure*}

Using the Geminid data from the night of 2015 December 14/15, we show the sensitivity of the flux to the elevation angle cutoff. The results, shown in Figure \ref{fig:ElevationAngleFlux}, demonstrate that the flux greatly decreases with decreasing elevation angle cutoff below 30 degrees. Such a trend strongly suggests that the observed sample of meteors is incomplete at low elevation angles. To avoid such incompleteness concerns, we restrict ourselves to an elevation angle cutoff of 30 degrees. 

\subsection{Luminous Efficiency}

The other primary source of systematic uncertainty in our mass index measurements involves the assumed model for the luminous efficiency of optical meteoroids. As discussed above, the peak magnitude - mass - velocity relationship that we adopt was ultimately derived from the results of \cite{Jacchia1967}, which assumed that the luminous efficiency was directly proportional to the meteor's speed \citep{Verniani1964}. The luminous efficiency of meteoroid ablation at optical wavelengths affects the mapping of limiting magnitudes to limiting masses, effectively shifting the limiting masses of the Wide-Field and All-Sky fluxes relative to the CMOR radar measurements. Literature models for the luminous efficiency vary by as much as an order-of-magnitude \citep[see e.g.][for a more detailed summary]{Subasinghe2017,Subasinghe2018}.

We have run an additional set of Monte Carlo simulations designed to account for potential systematic uncertainties in the luminous efficiency model used for both the Wide-Field and All-Sky limiting magnitude to mass conversion. For these simulations, we allowed the limiting masses to vary by $0.5 \thinspace \mathrm{dex}$ in either direction around the nominal value, corresponding to a factor of ten between the minimum and maximum limiting masses. Both the Wide-Field and All-Sky limiting masses were shifted by the same amount in each realization. We would expect ``coherent'' shifts for both data points under the assumption that a single luminous efficiency model affects both of these limiting masses. Tests on all five showers discussed in Section \ref{sec:ShowerResults} show only a small increase in the width of the resultant mass index distributions, with uncertainties increasing by $\sim 10-15\%$. The Orionids are most sensitive to this systematic uncertainty, since the mass index is measured using only CMOR and All-Sky measurements. The uncertainty in the Orionids mass index increases from $0.09$ to $0.10$, with the overall value remaining consistent. The Geminid data that we present remains practically unchanged even after accounting for potential uncertainty in the optical luminous efficiency. The presence of a flux measurement at an even higher mass limit from lunar impact monitoring ensures a stable mass index measurement. This relative stability even in the presence of large luminous efficiency uncertainties may appear surprising at first glance, but is largely driven by the fact that the spacing between the mass limits of the different instruments is significantly larger than the uncertainties on any particular mass limit. These large lever-arms enable a robust mass index measurement to be made even in the presence of large systematic uncertainties on any particular limiting mass.

\section{Comparisons to the MEO's Shower Forecast}

Many of the mass index measurements presented here are for showers whose mass index values are not well established by MEO observations. Instead, nominal values from the IMO or other sources are generally utilized for these showers when creating the annual meteor shower forecast \citep{Moorhead2017_Forecast}. We compare the three sets of mass indices in Table \ref{tab:IndexComps}. For four of the five showers, the measured mass indices are lower than those utilized in the shower forecast and by the IMO. For several of these showers (such as the Quadrantids and Orionids), the shower forecast mass index was taken directly from the IMO. As the discussion for each shower above should make clear, however, is that the IMO population indices do not account for the large variations in the measured values for each shower.

The forecasted flux on a given date to a given limiting mass depends on more variables than the mass index alone, however. In order to provide the most direct comparison between the model predictions and the observations, we utilize the MEO's past meteor shower forecasts to determine how well they predicted the flux on each of the five nights where shower fluxes were measured. These results are shown in Table \ref{tab:FluxComps}. The predicted fluxes for all five showers is consistently lower than the shower forecast prediction by $\sim 20-40\%$, although given the $\sim 25-50\%$ uncertainties on the flux measurements the the two values broadly agree. The flux measurements for the Leonid and Quadrantid showers, which are the two showers most discrepant with their forecasted values, occurred at dates significantly offset from the peak solar longitude assumed by the shower forecast. Since the assumed activity profile is assumed to follow a double-exponential form \citep{Moorhead2017_Forecast} around the peak, small uncertainties in the activity profile parameters can lead to larger uncertainties in the predicted flux at off-peak dates and times. Further investigations with larger numbers of shower fluxes at this mass limit will be required to identify the extent to which these discrepancies represent a systematic discrepancy between the observations and the shower forecast.


\begin{table*} 
    \centering
    \begin{tabular}{c c c c } \hline\hline
    Shower & IMO (r/s) & Forecast (r/s) & This Work (r/s) \\
    \hline
Geminids & 2.6/1.95 & 2.6/1.95 & 2.01/1.70 \\
Perseids & 2.4/1.87 & 2.2/1.79 & 1.72/1.54 \\
Leonids & 2.5/1.92 & 2.9/2.06 & 1.92/1.66 \\
Quadrantids & 2.1/1.74 & 2.1/1.74 & 2.30/1.83 \\
Orionids & 2.5/1.92 & 2.5/1.92 & 2.06/1.72\\

         \hline\hline
    \end{tabular}
    \caption{\label{tab:IndexComps} A comparison between the meteor shower mass and population indices measured in this work with those currently assumed by the IMO and what was assumed in the MEO's annual shower forecast for the year in question \citep{Moorhead2017_Forecast}.    }
\end{table*}

\begin{table*} 
    \centering
    \begin{tabular}{c c c  c } \hline\hline
    Shower &  Flux (Forecast) & Flux (This Work) & $B\vert \lambda_{\odot} - \lambda_{0} \vert$ \\
    \hline
Geminids &  $2.04 \times 10^{-5}$ & $1.64 \times 10^{-5}$ & 0.21\\
Perseids &  $4.86 \times 10^{-5}$ & $3.46 \times 10^{-5}$ & 0.09\\
Leonids &  $7.16 \times 10^{-6}$ & $4.85 \times 10^{-6}$ & 0.59\\
Quadrantids &  $1.08 \times 10^{-5}$ & $6.90 \times 10^{-6}$ & 0.85 \\
Orionids &  $4.85 \times 10^{-6}$ & $3.87 \times 10^{-6}$ & 0.07\\

         \hline\hline
    \end{tabular}
    \caption{\label{tab:FluxComps} A comparison between the meteor shower fluxes as predicted by the MEO's annual shower forecast and those measured in this work. For all fluxes, the units of flux are $\km^{-2} \hr^{-1}$. The fourth column denotes the difference between the solar longitude of the flux measurements and solar longitude of peak activity, normalized by the assumed characteristic scale time of the activity profile $B$ \citep[for more details see][]{Moorhead2017_Forecast}. Increasing values of $B\vert \lambda_{\odot} - \lambda_{0} \vert$ are sampling further into the ``tails'' of the assumed shower activity profile where the observations and model predictions become more uncertain.   }
\end{table*}

\section{Discussion and Future Prospects}
The process described here constitutes a new algorithm developed to measure calibrated fluxes (number of meteors per unit area and time down to a limiting mass) from All-Sky video cameras. The results from five showers demonstrate that these fluxes are fully consistent with expectations from independent observations in different mass regimes. No special ``fine-tuning" of parameters beyond the elevation angle cutoff were performed. The joint measurements of all detectors provide important constraints on the mass indices of meteor showers. We emphasize that the nominal mass limits of the All-Sky and Wide-Field camera networks span the majority of the masses that pose risk to spacecraft ($\sim 10^{-3} \g - 1 \g$). When combined with CMOR radar observations (with limiting masses of $\sim 10^{-4} \g$,) fluxes across the entirety of the spacecraft threat regime (in terms of meteoroid mass) can be inferred. 

The biggest limitation of this analysis procedure is the need to derive the limiting magnitude directly from the observed distribution of shower meteors from that night. While a method to determine the limiting meteor magnitude from the stars or some other independent source of measurements would be preferred, even small uncertainties in the selection function could lead to large uncertainties in the flux. The current method, while only feasible for the most active nights, is robust to systematic uncertainties since it already takes the selection function into account. The selection function is only taken into account on average across all cameras and times during the night, however, given the sample sizes needed for this analysis to be feasible. A larger number of cameras or a sample of more sensitive cameras may also help boost sample sizes. Given the realities of the observed sample sizes for these five showers, we do not expect there to be many more showers for which fluxes can be measured in this fashion with the All-Sky camera data. By design, however, these five showers are among the most active at masses of $\sim 1 \g$, and subsequently constitute the showers that pose the most risk to spacecraft.

In addition to requiring that the limiting magnitude of the shower sample be determined empirically, a large sample of shower meteors is also required to ensure that the surviving sub-sample is sufficiently large to provide meaningful results. For these five showers, approximately $50\%$ of the meteors are fainter than the limiting magnitude and are subsequently excluded from any flux calculation. Measuring the flux associated with events fainter than the limiting magnitude requires a full calculation of the selection function, however. 

The statistical nature of this procedure also hinders the time scales over which the flux can be measured. This current technique can currently only provide a single flux measurement for a given night. If there are prior reasons to suspect that the shower flux may change significantly over the course of a single night, then that temporal structure cannot be investigated using these algorithms. The practical implications of this limitation on spacecraft risk assessment will inevitably depend strongly on the particular spacecraft's operational constraints.

This work has also shown that, at least for these five showers, well calibrated visual meteor shower observations remain a powerful data source for the MEO. The excellent agreement between the equivalent ZHR for our All-Sky fluxes and measured visual ZHR for all five showers discussed here was essential to validating the general procedure and camera fluxes. Given the huge difference in limiting magnitude for the All-Sky camera fluxes and the visual observers\footnote{Approximately 9-10 magnitudes, corresponding to a factor of $\sim 10,000$ in luminosity.}, even small systematic errors in the mass/population index, limiting mass, or overall flux could result in largely discrepant extrapolated ZHR's. This procedure also suggests that a power-law distribution across this huge range of luminosities is a reasonable approximation to the shower mass distribution. Showers other than these five may not follow a power-law distribution over such a large dynamic range, however. Comparisons between visual ZHR measurements, Wide-Field fluxes, and All-Sky fluxes could be used, at least in principle, to identify showers where the mass distribution deviates from a power law over this mass range.

Although the mass index values for each of these five showers differs from what has been utilized in past meteor shower forecasts, the agreement between the predicted fluxes and measurements for these five showers are in very good agreement. Our analysis shows that while the All-Sky fluxes and mass indices measured here may improve future forecasts, accurate shower activity profiles can be just as important to predictions of shower fluxes at limiting masses of $\sim 1 \g$.  

Despite these limitations, the ability to measure fluxes and  mass indices from the All-Sky cameras greatly enhances their ability to inform the shower forecast and MEM. A joint All-Sky, Wide-Field, and CMOR flux analysis could be performed for every night with sufficient numbers of shower meteors, which would improve overall measurement precision and allow the MEO to monitor for variability in both activity and mass indices. All of these measurements would ultimately provide crucial new constraints and insights into the risk major meteor showers pose to spacecraft.   

\section*{Acknowledgements}

This work was supported by the NASA Meteoroid Environment Office under contract 80MSFC18C0011. We thank Dr. Bill Cooke for his support of this project, Dr. Althea Moorhead for providing assistance with the MEO's meteor shower forecast, and Aaron Kingery for providing the source code that became the clear time GUI. We also thank Dr. Peter Brown and Dr. Margaret Campbell-Brown at the University of Western Ontario for scrutinizing this process and providing important feedback on its viability. We finally thank the International Meteor Organization for providing a complete and easy-to-search database of visual meteor shower observations. These data proved crucial for validating the work presented here.

\bibliographystyle{elsarticle-harv}

\bibliography{AllRefs}

\end{document}